\documentclass[lettersize,journal]{IEEEtran}
\usepackage{amsmath,amssymb,amsfonts}

\usepackage{algorithmic}
\usepackage{algorithm}

\usepackage{graphicx}
\usepackage{textcomp}
\usepackage{xcolor}
\usepackage{pifont}
\usepackage[caption=false,font=normalsize,labelfont=sf,textfont=sf]{subfig}
\usepackage{dblfloatfix}

\usepackage{array}
\usepackage{multirow}
\usepackage{tabularx}
\usepackage{threeparttable}
\usepackage{booktabs}

\usepackage[T1]{fontenc}
\usepackage{verbatim}

\usepackage{url}
\usepackage{xurl}
\usepackage{cite}
\usepackage{hyperref}
\hypersetup{breaklinks=true}

\usepackage{enumitem}

\usepackage[most]{tcolorbox}

\usepackage{pgfplots}
\usepackage{pgfplotstable}
\pgfplotsset{compat=1.18}

\usepackage[mathlines,switch]{lineno}

\newcommand{\cmark}{\textcolor{green!60!black}{\ding{51}}}
\newcommand{\xmark}{\textcolor{red}{\ding{55}}}

\begin{document}

\title{Noisy Issues, Noisy Results: Mislabeling in Information Retrieval-based Bug Localization}

\author{Moumita Asad, Rafed Muhammad Yasir, Yirui He,
Ziyao He, and Sam Malek
\thanks{Moumita Asad, Rafed Muhammad Yasir, Yirui He,
Ziyao He, and Sam Malek are with the University of
California, Irvine, Irvine, CA, USA
(e-mail: \{moumitaa, ryasir, yiruih, ziyaoh5, malek\}@uci.edu).}}



\maketitle

\begin{abstract}
Information Retrieval-based Bug Localization (IRBL) leverages bug reports to identify faulty source files. Evaluations of IRBL approaches primarily rely on labeled issues from issue tracking systems. However, prior research has shown that issue labels are frequently incorrect in open-source projects, threatening the validity of IRBL evaluations. Existing studies have primarily examined traditional IRBL techniques on small datasets, overlooking modern approaches and the partial overlap between noisy and cleaned datasets. In this paper, we conduct a large-scale empirical study to investigate the prevalence and impact of mislabeled issues in IRBL benchmarks. We analyze five benchmark datasets comprising over 55,000 issue reports and find that, on average, 34.82\% of issues are mislabeled. We further evaluate six IRBL techniques, spanning traditional, deep learning-based, and large language model-based approaches using the widely adopted Ye et al. benchmark and statistical methods that account for dataset overlap. Our results show that removing mislabeled issues from evaluation datasets significantly reduces measured performance across all six techniques, with Mean Reciprocal Rank decreasing by 4.68\%–21.74\%. Although mislabeling does not change the relative ranking of techniques, it affects the magnitude of their performance differences. These findings highlight the importance of reliable evaluation data for ensuring valid and meaningful comparisons of IRBL techniques.
\end{abstract}


\section{Introduction}
Issue reports are critical artifacts in software engineering projects, serving as documented records of bugs, enhancement requests, and other development tasks \cite{kochhar2014potential}. These reports are generally managed through Issue Tracking Systems (ITSs), where users and contributors can submit, track, and resolve issues. Among various types of issue reports, bug reports are particularly important as unresolved bugs can compromise the reliability, performance and security of software systems \cite{kochhar2014s}. However, modern software projects often receive a large volume of bug reports, making it challenging for developers to manage them. For example, the Eclipse project received over 4,400 bug reports in 2009 alone \cite{zhou2012should}. Addressing each report requires developers to reproduce the bug, identify its root cause and locate the relevant source code entities, a process that is both time-consuming and labor-intensive \cite{7390282}.

To assist developers in this process, various Information Retrieval-based Bug Localization (IRBL) techniques have been proposed. These approaches formulate bug localization as a document retrieval task, where the bug report serves as a query and the source files as documents. Given a bug report, these techniques generate a ranked list of source files based on their likelihood of containing the fault \cite{qi2021dreamloc}. The goal is to rank the buggy files as high as possible. To calculate the similarity between bug reports and the source files, existing techniques use approaches ranging from traditional models such as the Vector Space Model (VSM) \cite{rao2011retrieval} and Topic Modeling \cite{lukins2010bug,nguyen2011topic} to advanced methods based on Deep Learning \cite{lam2015combining,lam2017bug,xiao2019improving,zhang2020exploiting,qi2021dreamloc,liang2022modeling,chakraborty2024rlocator} and Large Language Models (LLMs) \cite{Brain2025,li2025llm,asad2026towards}.

To evaluate the effectiveness of these techniques, several benchmark datasets have been constructed from ITSs. These datasets generally collect fixed issues labeled "bugs", assuming that these labels are correct \cite{lee2018bench4bl}. However, a study by Herzig et al. found that approximately 33.8\% of the bug reports are misclassified (i.e., these issues do not actually describe bugs) \cite{herzig2013s}. This finding raises concerns about the quality of IRBL benchmark datasets and the validity of performance evaluations conducted using them. Since IRBL techniques are designed for bug reports, the inclusion of non-bug issues may introduce bias into their evaluation \cite{kochhar2014potential}.

Despite the importance of this problem, only a few studies have examined the impact of mislabeled issues on IRBL performance. Kochhar et al. first examined this problem using VSM-based bug localization \cite{kochhar2014s,kochhar2014potential}, while Widyasari et al. subsequently investigated two traditional approaches, VSM and BugLocator \cite{widyasari2022influence}. However, as summarized in Table \ref{diff}, these studies have several limitations. First, they focused on traditional IRBL techniques, leaving the impact of mislabeling on more advanced deep learning and LLM-based approaches (e.g., DreamLoc\cite{qi2021dreamloc}, GenLoc \cite{asad2026towards}) largely unexplored. Second, their analyses were conducted on small datasets and did not extend to established IRBL benchmark datasets (e.g., Ye et al. dataset \cite{ye2015mapping}, Bench4BL \cite{lee2018bench4bl}). Third, their evaluations were limited to intra-technique comparisons (i.e., analyzing how the performance of individual techniques changes between noisy and clean datasets) without performing inter-technique comparisons, which assess how mislabeled issues affect the comparative performance of different techniques. Finally, they overlooked the partially overlapping nature of the data in which the clean dataset is a subset of the original noisy dataset. Consequently, they applied statistical tests that assume data independence, which may lead to misleading conclusions. As a result, the validity of prior findings remains uncertain.

% \begin{table}[h]
% \centering
% \caption{Comparison with prior studies investigating mislabeled issues in IRBL.}
% \label{tab:prior-comparison}
% \small
% \resizebox{\columnwidth}{!}{
% \begin{tabular}{lccccc}
% \toprule
% \textbf{Study} &
% \begin{tabular}[c]{@{}c@{}}\textbf{Established IRBL}\\\textbf{Benchmarks}\end{tabular} &
% \begin{tabular}[c]{@{}c@{}}\textbf{DL/LLM}\\\textbf{Techniques}\end{tabular} &
% \begin{tabular}[c]{@{}c@{}}\textbf{Intra-Technique}\\\textbf{Comparison}\end{tabular} &
% \begin{tabular}[c]{@{}c@{}}\textbf{Inter-Technique}\\\textbf{Comparison}\end{tabular} &
% \begin{tabular}[c]{@{}c@{}}\textbf{Partial-Overlap}\\\textbf{Aware Analysis}\end{tabular} \\
% \midrule
% Kochhar et al.      & \xmark & \xmark & \cmark & \xmark & \xmark \\
% Widyasari et al.    & \xmark & \xmark & \cmark & \xmark & \xmark \\
% \textbf{Our Study}  & \cmark & \cmark & \cmark & \cmark & \cmark \\
% \bottomrule
% \end{tabular}
% }
% \label{diff}
% \end{table}

\begin{table}[h]
\centering
\caption{Comparison with prior studies investigating mislabeled issues in IRBL.}
\label{tab:prior-comparison}
\small
\resizebox{\columnwidth}{!}{
\begin{tabular}{lccc}
\toprule
\textbf{Comparison Criteria} &
\textbf{Kochhar et al.} &
\textbf{Widyasari et al.} &
\textbf{Our Study} \\
\midrule
Established IRBL Benchmarks
    & \xmark & \xmark & \cmark \\
DL/LLM Techniques
    & \xmark & \xmark & \cmark \\
Intra-Technique Comparison
    & \cmark & \cmark & \cmark \\
Inter-Technique Comparison
    & \xmark & \xmark & \cmark \\
Partial-Overlap Aware Analysis
    & \xmark & \xmark & \cmark \\
\bottomrule
\end{tabular}
}
\label{diff}
\end{table}

In this study, we address these limitations through a comprehensive empirical investigation of the prevalence of mislabeled issues and their impact on IRBL performance. We extend prior work by examining established IRBL benchmarks and modern localization techniques, evaluating the effects of mislabeling on both individual and comparative performance, and employing statistical methods that account for dataset overlap. In particular, we investigate the following two research questions: 
\vspace{-4pt}
\begin{description}[leftmargin=*]
    \item \textbf{RQ1.} How prevalent are mislabeled issues in IRBL benchmark datasets?
    \item \textbf{RQ2.} How do mislabeled issues impact the performance of IRBL techniques?
\end{description}

To address RQ1 (Section \ref{sec:rq1}), we investigate five IRBL benchmark datasets, including three widely used datasets (Zhou et al. \cite{zhou2012should}, Ye et al. \cite{ye2015mapping} and Bench4BL \cite{lee2018bench4bl}) and two recently introduced datasets (GHRB (GitHub Recent Bugs) \cite{lee2024github} and BeetleBox \cite{chakraborty2025blaze}). Together, these benchmarks contain more than 55,000 issue reports spanning over a decade of IRBL research. To identify mislabeled issues at scale, we develop GenTag, an LLM-based issue classifier that achieves a weighted F1-score of 90\% while maintaining an affordable cost of \$0.0003 per issue. Applying GenTag to these datasets reveals that mislabeled reports are widespread, accounting for 29.01\% to 44.01\% of the issues across the five benchmarks, with an average of 34.82\%, highlighting persistent limitations in benchmark label quality.

To address RQ2 (Section \ref{sec:rq2}), we evaluate six representative IRBL techniques, namely VSM \cite{kochhar2014potential}, BugLocator \cite{zhou2012should}, BLUiR \cite{saha2013improving}, BRTracer \cite{wong2014boosting}, DreamLoc \cite{qi2021dreamloc}, and GenLoc \cite{asad2026towards}. These techniques span traditional retrieval, heuristic-enhanced, deep learning-based, and LLM-based approaches. Using the widely adopted Ye et al. dataset \cite{ye2015mapping}, we investigate the effects of mislabeled issues in both evaluation and historical/training data. We conduct intra-technique comparisons to assess how mislabeling affects individual techniques and inter-technique comparisons to examine its influence on their relative performance. Our statistical analyses account for the overlapping nature of the evaluated datasets. Our results show that removing mislabeled issues from the evaluation data consistently reduces the measured performance of all six techniques, with statistically significant differences. In contrast, cleaning historical/training data has varying effects depending on the technique. Furthermore, although mislabeling does not change the overall ordering of techniques, it influences the magnitude of performance differences between them. These findings highlight the importance of dataset quality for fair and accurate evaluation in IRBL research.

In summary, this paper makes the following contributions: 
\begin{itemize}
\item \textbf{Large-scale analysis of issue mislabeling:} We investigate the prevalence of issue mislabeling across five IRBL benchmarks comprising over 55,000 issue reports using a validated LLM-based classifier.
\item \textbf{Impact on intra-technique comparisons:} We systematically evaluate the effects of mislabeling in evaluation and historical/training data on the performance of six IRBL techniques, spanning traditional, deep learning-based, and LLM-based approaches.
\item \textbf{Impact on inter-technique comparisons:} We examine how issue mislabeling affects the relative performance and statistical comparisons among IRBL techniques.
\item \textbf{Replication package:} We release a replication package \cite{repo} to facilitate reproducibility and support future research.
\end{itemize}

\section{Prevalence of Mislabeled Issues in IRBL Datasets}
\label{sec:rq1}

\subsection{Identification of Mislabeled Issues}

To assess the prevalence of mislabeled issues in IRBL datasets, an automated classifier is required to distinguish between bug and non-bug reports, as manual labeling is time-consuming \cite{santos2023tag} and does not scale to large benchmarks (e.g., over 20,000 reports) \cite{herzig2013s}. Existing approaches based on machine learning, deep learning or BERT-like models (e.g., RoBERTa, SetFit) \cite{pingclasai2013classifying,pandey2017automated,kallis2019ticket,herbold2020feasibility,li2022deeplabel,colavito2023few} rely on manually verified data for training or fine-tuning, which is often unavailable and therefore limits their applicability. Recently, LLM-based classifiers \cite{colavito2024leveraging,colavito2025benchmarking} have been introduced to reduce this dependency. However, these models are generally provided with context about only a limited set of categories (e.g., \texttt{BUG}, \texttt{FEATURE}, \texttt{DOCUMENTATION}, \texttt{QUESTION}) and prompted to choose among them. When the true label is absent from the prompt, the LLM is forced to map the issue to one of the available labels. For example, issues related to \texttt{REFACTORING} or \texttt{BUILD\_SYSTEM} could be misclassified as \texttt{BUG}, which could inflate bug counts and bias subsequent analyses. This limitation underscores the need for incorporating a broader taxonomy into LLM prompts to enable more accurate and fine-grained issue classification.

Motivated by this limitation, we design \textbf{GenTag} (\textbf{Gen}erative AI-based Issue \textbf{Tag}ger), an LLM-powered classifier designed to perform fine-grained issue classification \footnote{GenTag is a supporting tool for the empirical study and is not the primary contribution of this paper.}. GenTag leverages the taxonomy and guidelines proposed by Herzig et al., which define 13 comprehensive issue categories such as \texttt{BUG}, \texttt{RFE} (Request for Enhancement), and \texttt{REFACTORING} \cite{herzig2013s}. Based on these guidelines, GenTag instructs an LLM to classify each issue report into one of the predefined issue categories.

The classification process begins with a prompt that tells an LLM to act as an expert software maintainer, as role-playing strategies have been shown to improve the reasoning skills of LLMs \cite{kong2023better}. The prompt then includes the 13 issue categories along with their descriptions. However, since the original Herzig et al. guidelines were designed for human annotators, we refined the guidelines through prompt engineering to make them more suitable for LLM-based classification. We applied two types of modifications:

\begin{itemize}[leftmargin=*]
\item \textbf{Text Reduction:} To reduce token usage and simplify the prompt, we condensed complex definitions and removed redundant ones. Specifically, we shortened the description of the \texttt{IMPROVEMENT} category by consolidating Herzig et al.'s detailed nine cases into four broader subcategories, focusing on feature enhancements and error handling, exceptions, performance or compatibility improvements. Additionally, we removed definitions for categories such as \texttt{BACKPORT} (backports), \texttt{CLEANUP} (code cleanups), \texttt{SPEC} (changes to specification rather than documentation or code) and \texttt{TASK} (general development tasks), as we found that these definitions provided little added value for LLM classification. These adjustments reduced token consumption and inference cost without degrading classification performance.

\item \textbf{Text Enhancement:}
To improve the precision of the classification, we clarify certain category definitions. For \texttt{BUILD\_SYSTEM}, we added explicit examples of configuration files (e.g., pom.xml, build.xml) to help identify them better. We also revised the vague definition of the \texttt{TEST} category, which originally covered all "issues regarding test cases". Our refined definition explicitly included issues related to test coverage, testing frameworks, test case initialization, as well as typos or improvements in test methods or test logic. Additionally, we clarified to the LLM that test-related changes addressing broken or malfunctioning tests intended to restore correct behavior should be classified as bugs, not as test issues. This distinction helped to ensure that bug-fixing changes in the test code were correctly labeled as \texttt{BUG} rather than mistakenly categorized under \texttt{TEST}. These enhancements provided clearer guidance and improved the LLM’s classification performance.
\end{itemize}

The prompt then tells the LLM to classify an issue in the most appropriate category by providing the title and description of the issue. The prompt additionally asks the LLM to provide a brief explanation of its decision, as previous work has shown that self-explaining prompts significantly improve LLM comprehension \cite{wu2023large}. The complete prompt used by GenTag is provided in our replication package \cite{repo}.

While GenTag performs fine-grained classification into 13 categories, our ultimate goal is to analyze the impact of mislabeled issues on bug localization. Since bug localization techniques operate on a binary distinction between bug-related and non-bug-related reports, we map the fine-grained categories into two classes: \texttt{BUG} and \texttt{NOT BUG}. The \texttt{BUG} category remains unchanged, whereas all remaining categories (e.g. \texttt{RFE}, \texttt{REFACTORING}, and \texttt{DOCUMENTATION}) are labeled as \texttt{NOT BUG}. This abstraction allows us to preserve the benefits of fine-grained annotation while aligning the final labels with the requirements of bug localization evaluation.

Since LLM outputs are inherently non-deterministic, we repeat the classification process three times and select the final label through majority voting. This approach mitigates potential variability in LLM predictions and ensures more stable results \cite{chen2024more}.

% \enlargethispage{2\baselineskip}
% \vspace{-10pt}
\subsection{Validation of the Identification Process} 

\noindent \textit{Baseline Techniques, Dataset and Evaluation Metrics:} To evaluate the performance of GenTag, we compared it against a state-of-the-art LLM-based approach, called \textit{GPT-like model}, that classifies an issue into categories such as bug, feature, documentation and question under a zero-shot setting \cite{colavito2024leveraging}. In addition, we evaluated a variant of GenTag (GenTag-Herzig) that directly follows the original issue classification guidelines proposed by Herzig et al. to assess the impact of prompt engineering.

For evaluation, we used Herzig et al.'s dataset \cite{herzig2013s}, which consists of 5,591 hand-labeled issue reports from three projects (HTTPClient, Jackrabbit and Lucene-Java). Classification performance was measured using standard metrics, namely precision, recall and F1-score. Following prior work \cite{li2022deeplabel}, we report both class-specific scores for the BUG and NOT BUG categories as well as weighted averages to account for class imbalance. The weighted metrics are computed using the proportion of instances belonging to each class.

\noindent \textit{Experiment Setup:}
GenTag uses \texttt{GPT‑4o mini} by OpenAI due to its cost-effectiveness (\$0.15 per 1M input tokens and \$0.60 per 1M output tokens) \cite{gpt-mini}. We set the model’s temperature to 0 to reduce variability and facilitate reproducibility \cite{nashid2023retrieval}. For comparison, we adapted the replication package of the \textit{GPT-like model} to support Herzig's dataset and used the same underlying LLM as GenTag to ensure a fair comparison. Following the original study \cite{colavito2024leveraging}, we executed the GPT-like model three times to account for the non-deterministic nature of LLMs and reported the average results.

\noindent \textit{Result:}
Table \ref{performance} compares the performance of GenTag against two baselines: the GPT-like model \cite{colavito2024leveraging}, a state-of-the-art LLM-based issue classifier, and GenTag-Herzig, a variant of GenTag that adheres strictly to Herzig et al.'s original classification guidelines. Since GenTag serves as a supporting tool rather than the primary focus of this paper, we do not present the detailed results of each individual classification trial in the manuscript due to space limitations. The results of all three GenTag trials, together with the agreement analysis across trials, are available in the replication package.

% \vspace{2pt}
\begin{table}[h]
\centering
\caption{Performance of Issue Classification Approaches}
\tabcolsep=0.11cm
\centering
% \scriptsize
\begin{tabular}{|l|ccc|ccc|ccc|}
\hline
\multirow{2}{*}{\textbf{Approach}}                                        & \multicolumn{3}{c|}{\textbf{Bug}}             & \multicolumn{3}{c|}{\textbf{Not-bug}}         & \multicolumn{3}{c|}{\textbf{Weighted}}          \\ \cline{2-10} 
                                                         & \textit{prec} & \textit{rec}  & F1            & \textit{prec} & \textit{rec}  & F1            & \textit{prec} & \textit{rec}  & F1 \\ \hline

\begin{tabular}[c]{@{}l@{}}GPT-like\\ model\end{tabular} & 0.61          & \textbf{0.98} & 0.76          & \textbf{0.99} & 0.67          & 0.80          & 0.86          & 0.78          & 0.78            \\ \hline
\begin{tabular}[c]{@{}l@{}}GenTag-\\Herzig\end{tabular}              & \textbf{0.91} & 0.65          & 0.76 & 0.84          & 0.96          & 0.90          & 0.86         & 0.85          &        0.85   \\ \hline
\textbf{GenTag}                                          & 0.83          & 0.88          & \textbf{0.86} & 0.94          & 0.90          & \textbf{0.92} & \textbf{0.90} & \textbf{0.90} & \textbf{0.90}   \\ \hline
\end{tabular}
\label{performance}
\end{table}
% \vspace{2pt}

Results show that GenTag obtains the highest F1-score for both bug (F1 = 0.86) and not-bug (F1 = 0.92) issues as well as the highest weighted metrics (precision = 0.90, recall = 0.90 and F1 = 0.90). In contrast, the GPT-like model baseline achieves exceptionally high precision on the not-bug class (0.99), but this comes at the expense of recall (0.67), resulting in a lower F1-score (0.80). Similarly, GenTag-Herzig attains high precision on the bug class (0.91), yet its lower recall (0.65) reduces overall effectiveness (0.76). These findings highlight the importance of maintaining balanced precision and recall across classes, a balance that GenTag successfully achieves without compromising either metric.

% Overall, \textbf{GenTag} delivers the most balanced performance, achieving the highest F1-scores for both the bug (0.86) and not-bug (0.92) classes, as well as the best weighted results (precision = 0.90, recall = 0.90, F1 = 0.90).

% These trade-offs underscore the importance of balanced performance across classes—something that GenTag consistently achieves without sacrificing one metric for another.

% Although the GPT-Like model achieves exceptionally high recall on the bug class (0.98), but this comes at the expense of precision (0.61), resulting in a lower F1-score (0.76). Similarly, GenTag-Herzig attain strong not-bug precision (0.99), yet their weaker not-bug recall (0.67)  drags down their overall effectiveness. These trade-offs highlight the importance of balanced performance across classes--something that GenTag consistently achieves without compromising one metric for another. 

% A sample case is shown in Figure \ref{jcr_issue}, where GenTag correctly classified the issue as a refactoring issue, while GPT-like model misclassified it as a bug. This highlights how GenTag’s fine-grained classification and refined prompts allow it to classify issues more accurately. Since GPT-like model lack certain issue categories, it assumes issues like refactoring as bugs. 

A sample case is shown in Fig.~\ref{jcr_issue}, which describes a simple spelling mistake (“Trasaction” vs. “Transaction”). Since this issue does not cause runtime or memory errors or require semantic changes to fix faulty functionality, it is categorized as \texttt{REFACTORING} during manual labeling. In this case, GenTag correctly classified the issue as refactoring, whereas the GPT-like model baseline misclassified it as a bug due to its restricted label space which does not include a \texttt{REFACTORING} category. This example shows how GenTag’s fine-grained taxonomy helps prevent issues from being mislabeled as bugs, thereby yielding more accurate classifications for downstream IRBL studies.

\begin{figure}[htb]
\centering
\begin{tcolorbox}[
  % enhanced,
  % parbox=false,
  % before upper=\strut,
  top=2pt,
  bottom=2pt,
  left=2pt,
  right=2pt,
  % boxsep=4pt,
  colback=gray!10!white, 
  colframe=black!75!black, 
  % boxrule=0.5pt, 
  % arc=2mm,
  % width=0.98\textwidth,
  % toprule at break=0mm,
  % bottomrule at break=0mm,
]
% \small
\raggedright
\fontsize{7.8pt}{7.8pt}\selectfont
% \textbf{Bug ID:} JCR-667 \\
\textbf{Summary:} Variant spelling "Trasaction" and "Transaction" now in codebase\\
\textbf{Description:} Incorrectly spelled "Trasaction" now in codebase. The correctly spelled "Transaction" appears a far greater number of times. Isolated to package: org.apache.jackrabbit.jca \\
Incorrectly  spelled "Trasaction" found in: \\
- (3x) JCAManagedConnection.java \\
- (7x) JCAManagedConnectionFActory.java \\
Correctly spelled "Transaction" found in: \\
- (20x) JCAManagedConnection.java \\
- (2x) JCAManagedConnectionFActory.java \\
- (1x) JCAResourceAdapter.java \\
- (12x) TransactionBoundXAResource.java 
\end{tcolorbox}
\caption{Bug Report (ID: JCR-667) from the JCR Project}
\label{jcr_issue}
\end{figure}

In addition to achieving strong performance, GenTag is highly cost-effective. Running it on 5,591 issues from Herzig et al.'s dataset costs only \$1.56 (approximately \$0.0003 per issue), making it feasible to apply to large-scale IRBL datasets.

% GenTag is also highly cost-effective. Running it three times on Herzig et al.'s dataset costs only \$1.56, demonstrating its affordability scale. This low cost makes it feasible to apply GenTag across multiple datasets, making it a practical option for large-scale preprocessing and evaluation workflows.

\subsection{Analysis of IRBL Benchmark Datasets}

To assess the prevalence of mislabeled issues, we applied GenTag to five established IRBL datasets. These include three widely used datasets: Zhou et al. \cite{zhou2012should}, containing 3,479 bug reports from four open-source Java projects; 
Ye et al. \cite{ye2015mapping}, comprising over 22,000 bug reports collected from six large open-source Java projects; 
and Bench4BL \cite{lee2018bench4bl}, containing 10,019 bug reports from 51 open-source Java projects. 
We also included two recently introduced datasets: GHRB (GitHub Recent Bugs) \cite{lee2024github}, containing 131 recent bug reports from 16 Java projects; 
and BeetleBox \cite{chakraborty2025blaze}, a large multi-language dataset comprising 19,146 bug reports from 29 open-source projects across five programming languages (Java, C++, Python, Go and JavaScript) \footnote{The original BeetleBox dataset contains 23,782 issues. Duplicate reports were removed prior to analysis, resulting in 19,146 unique bug reports.}.

To further validate GenTag, we manually labeled a stratified random sample of 382 issue reports drawn from these five datasets. The sample size corresponds to a 95\% confidence level and a 5\% margin of error, and the number of sampled issues from each dataset was proportional to its share of the combined corpus. Two authors independently categorized these issues based on Herzig et al.’s guidelines \cite{herzig2013s} and resolved discrepancies through discussion to reach a consensus. Inter-rater reliability, measured by Krippendorff’s Alpha, was 0.842, which indicates strong agreement among raters \cite{marzi2024k}. The consensus labels were then compared with GenTag’s predictions to assess its performance. 

\begin{table}[h]
\centering
\caption{Performance of GenTag on Hand Labeled Issues}
\begin{threeparttable}
\begin{tabular}{|l|r|r|r|r|}
\hline
\textbf{Class}     & \multicolumn{1}{l|}{\textbf{Precision}} & \multicolumn{1}{l|}{\textbf{Recall}} & \multicolumn{1}{l|}{\textbf{F1-Score}} & \multicolumn{1}{l|}{\textbf{Support}} \\ \hline
Bug                & 0.96                                    & 0.93                                 & 0.95                                   & 262                                    \\ \hline
Not Bug            & 0.85                                    & 0.93                                 & 0.89                                   & 120                                    \\ \hline
Weighted & 0.93                                    & 0.93                                 & 0.93                                   & 382                \\ \hline
\end{tabular}

\begin{tablenotes}[flushleft]
\footnotesize
\setlist[itemize]{leftmargin=*}
\item Support denotes the number of ground-truth instances for that class.
\end{tablenotes}
\end{threeparttable}
\label{tab:manual_classify_results}
\end{table}

Table~\ref{tab:manual_classify_results} presents the aggregated performance of GenTag on the manually labeled issues. Dataset-specific precision, recall and F1-scores are available in the replication package. GenTag achieves a weighted F1-score of 0.93, demonstrating strong generalization beyond its original evaluation dataset. These results confirm that GenTag can be reliably applied to IRBL datasets.

\subsection{Prevalence of Mislabeled Issues}

Table~\ref{fl-datasets} reveals a consistent presence of mislabeled issues across all IRBL datasets. Across the five datasets, the proportion of non-bug reports ranges from 29.01\% to 44.01\%, with an average of 34.82\%, indicating that roughly one-third of the issues labeled as bugs are actually non-bug reports. This finding aligns with the prior study by Herzig et al. \cite{herzig2013s}, which reported a similar proportion of mislabeled issues; however, their analysis focused on several open-source projects rather than IRBL benchmark datasets. Our analysis extends this observation by showing that IRBL datasets proposed both before and after the study of Herzig et al. are affected by mislabeled issues. Notably, even recently introduced benchmarks such as GHRB and BeetleBox contain a substantial proportion of non-bug reports. This pattern across datasets spanning more than a decade highlights that the assumption of clean ground truth in IRBL datasets does not always hold, and such noise has the potential to skew the evaluation of localization techniques. Therefore, it is important to examine whether dataset cleaning is necessary, depending on the extent to which mislabeled issues impact the evaluation of IRBL techniques.

\vspace{-2pt}
\begin{table}[h]
\caption{Mislabeled Issue Reports in IRBL Datasets}
\centering
\tabcolsep=0.12cm
\begin{tabular}{|l|c|r|r|r|}
\hline
\textbf{Dataset} & \textbf{Year} & \multicolumn{1}{c|}{\textbf{\begin{tabular}[c]{@{}c@{}}No. of\\ Issues\end{tabular}}} &
\multicolumn{1}{c|}{\textbf{\begin{tabular}[c]{@{}c@{}}Bug \\ Reports (\%)\end{tabular}}} & \multicolumn{1}{c|}{\textbf{\begin{tabular}[c]{@{}c@{}}Non-bug\\ Reports (\%)\end{tabular}}} \\ 
\hline
Zhou et al.    & 2012 & 3,479  & 55.99\% & 44.01\% \\ \hline
Ye et al.      & 2015 & 22,747 & 65.42\% & 34.58\% \\ \hline
Bench4BL       & 2018 & 10,019 & 69.29\% & 30.71\% \\ \hline
GHRB           & 2024 & 131    & 70.99\% & 29.01\% \\ \hline
BeetleBox      & 2025 & 19,146 & 64.21\% & 35.79\% \\ \hline
\end{tabular}
\label{fl-datasets}
\end{table}

\section{Impact of Mislabeled Issues on Performance of IRBL Techniques}\label{sec:rq2}

\subsection{IRBL Techniques}
To investigate the impact of mislabeled issues, we first defined selection criteria to guide the choice of IRBL techniques. Our goal is to ensure that the selected approaches are reproducible, scalable and aligned with the scope of our study. In particular, we considered techniques that (1) have a publicly available and functional implementation, (2) complete the localization of each bug report within 80 seconds, as followed in prior studies \cite{le2015information,arcega2021bug}, (3) perform file-level localization, and (4) support Java projects, since our benchmarks consist of large-scale Java systems and the majority of IRBL techniques and datasets in the literature are Java-based \cite{ye2015mapping,qi2021dreamloc,chakraborty2024rlocator}.

Based on these criteria, we selected six representative IRBL techniques that cover different methodological paradigms, including traditional information retrieval model, heuristic-enhanced approach, deep learning-based technique, and recent LLM-driven method. A brief overview of each technique is provided below:

\textbf{Vector Space Model (VSM):} It is one of the earliest IRBL approaches and has served as the basis for many state-of-the-art techniques \cite{kochhar2014potential}. In VSM, the bug report and the source files are represented as vectors of weighted terms, where the weights are computed using the Term Frequency-Inverse Document Frequency (TF-IDF) scheme. Next, the similarity between the bug report and each source file is measured using cosine similarity.

\textbf{BugLocator:} It extends VSM by using a revised Vector Space Model (rVSM) that accounts for file length, assigning higher scores to longer files since they are more bug-prone \cite{zhou2012should}. Additionally, it leverages historical bug data by identifying previously fixed bugs similar to the current report and boosts the rank of files involved in those fixes (SimiScore). Finally, files are ranked by combining the rVSM-based similarity and the SimiScore using a weighted sum.

\textbf{BLUiR:} Building on the limitations of earlier models, BLUiR enhances bug localization by leveraging the structure of source code \cite{saha2013improving}. It extracts elements like class names, method names, and comments from each file. Bug reports are similarly processed and treated as structured queries. Using a BM25-based retrieval model, BLUiR computes relevance scores by matching bug report sections with corresponding code elements. Lastly, files are ranked based on aggregated scores from all field combinations, giving more weight to terms appearing in multiple regions.

\textbf{BRTracer:} It integrates two key heuristics for bug localization: segmentation and stack-trace analysis \cite{wong2014boosting}. To reduce noise from large files, BRTracer splits each source file into segments and selects the segment most similar to the bug report to represent the file. Additionally, it analyzes stack traces in bug reports to identify directly mentioned files and their dependencies, assigning boost scores based on their presence and position in the trace. These enhancements are combined with a revised VSM-based score to produce a final ranking.

\textbf{DreamLoc:} It is a deep learning-based IRBL framework that uses a Wide and Deep architecture \cite{qi2021dreamloc}. The Wide component is a linear model that combines software-specific features such as VSM similarity, cyclomatic complexity, and bug-fix history. The Deep component uses a fine-grained relevance matching model, leveraging word embeddings and an attention mechanism to compare bug reports with tokenized code snippets. A fusion layer combines both components to produce a final score for each file.

\textbf{GenLoc:} GenLoc combines semantic retrieval with LLM-driven iterative code exploration to identify buggy files \cite{asad2026towards}. At first, it retrieves semantically relevant files for a given bug report using embedding-based similarity. Next, it leverages an LLM to analyze a bug report and further inspect these retrieved files, or explore additional files as needed. To enable this exploration, GenLoc provides the LLM with a set of functions (e.g., retrieving method signatures or method bodies) that support iterative analysis about which files are likely related to the reported bug.

Collectively, these techniques reflect the evolution of IRBL research. VSM and BugLocator represent foundational models; BLUiR and BRTracer incorporate structural and heuristic enhancements; DreamLoc exemplifies deep learning-based solutions; and GenLoc reflects the emerging paradigm of LLM-based bug localization. This diversity allows us to comprehensively assess how mislabeled issues affect techniques with varying dependencies on textual content, structural information and learning mechanisms. 

For DreamLoc and GenLoc, we used the original implementations. For BugLocator, BLUiR, and BRTracer, and VSM, we used the implementations provided by \cite{lee2018bench4bl} and \cite{widyasari2022influence}, respectively. Several other IRBL techniques (e.g., \cite{xiao2019improving,chakraborty2024rlocator,Brain2025}) were excluded because they lacked publicly available implementations, had unresolved dependency issues, or exhibited scalability limitations. Detailed inclusion and exclusion decisions are provided in the replication package \cite{repo}.

\subsection{Dataset, Configurations, and Evaluation Metrics}\label{sec:dataset}
We evaluated the six techniques on the Ye et al. dataset \cite{ye2015mapping}, which remains the most widely used benchmark in IRBL research \cite{niu2025deep} and continues to serve as the evaluation dataset for recent techniques \cite{chakraborty2025blaze,asad2026towards,malhotra2025validating}. Moreover, the size of the dataset (over 22,000 bug reports) makes it suitable for training DreamLoc, a deep learning–based approach. The other datasets used in RQ1 are less suitable for evaluating techniques that rely on training (i.e., DreamLoc) or historical bug-fix data (i.e., BugLocator, BRTracer). The GHRB dataset is relatively small, while the Bench4BL and Zhou et al. datasets contain several projects with fewer than 100 bug reports, making it difficult to obtain sufficient project-level training or historical data. Similarly, BeetleBox is primarily designed for cross-project bug localization scenarios (i.e., training on one project and testing on different projects), which does not align with the design of these approaches \cite{chakraborty2025blaze}. Hence, the Ye et al. dataset provides the most appropriate setting for evaluating the impact of mislabeled issues on all prior techniques. 

Following prior work \cite{han2023bjxnet,chakraborty2024rlocator}, we chronologically divided the original dataset into 60\% historical/training data and 40\% test data based on issue creation time. Earlier reports were used as historical or training information, while later reports were reserved for evaluation to avoid future-information leakage \cite{ye2015mapping,fejzer2021tracking}.

We prepared three configurations of the Ye et al. dataset to systematically investigate the impact of mislabeled issues in evaluation and historical/training data: \textbf{Original (ORIG), Test-Cleaned (TC), and Fully-Cleaned (FC)}. In the ORIG configuration, both historical/training and test partitions retain all issue reports, including those identified as non-bugs by GenTag. In the TC configuration, non-bug reports are removed only from the test partition, while the historical/training partition remains unchanged. Finally, in the FC configuration, non-bug reports are removed from both partitions. The same chronological split was maintained across all three configurations, with non-bug reports removed from the corresponding partitions according to each configuration. Thus, TC and FC share the same set of bug reports for evaluation but differ in the historical/training information available to the techniques.

These configurations enable us to separately examine two sources of dataset contamination. First, we compare ORIG and TC to assess how mislabeled reports in the test set influence measured localization performance. This comparison is conducted for all six IRBL techniques. Second, we compare TC and FC to investigate the impact of mislabeled reports in historical/training data while keeping the evaluation set fixed. This comparison applies only to techniques that leverage historical bug-fix information or require training data, namely BugLocator, BRTracer, and DreamLoc. For VSM, BLUiR, and GenLoc, which do not use such information, cleaning the historical/training partition has no effect on their inputs. Therefore, we exclude these techniques from the TC–FC comparison.

Each technique was evaluated under its applicable configurations. For GenLoc, we repeated the experiments three times to account for the inherent variability of LLM outputs. In addition to these intra-technique comparisons, we performed inter-technique comparisons to investigate how mislabeled issues affect the relative performance of different IRBL approaches. Specifically, we compared pairs of techniques on ORIG and TC using the same evaluation reports within each configuration. We used the following two evaluation metrics to assess localization performance across these comparisons:

% \begin{enumerate}[leftmargin=*]
\textbf{Accuracy@k:} It measures the percentage of bugs for which at least one of the buggy files is ranked within the top-k positions (k = 1, 5, 10) of the ranked list \cite{wong2023handbook}. A higher Accuracy@k value indicates a better performance of the bug localization technique. Similar to other studies \cite{rahman2018improving,qi2021dreamloc,liang2022modeling}, k is set to 1, 5 and 10.

 \textbf{Mean Reciprocal Rank (MRR):} This metric calculates the average reciprocal rank of the first correctly identified buggy file across all bug reports \cite{wong2023handbook}. For each bug report, the reciprocal rank is defined as the inverse rank of the first correct buggy file. MRR is calculated as follows:
\begin{equation}
MRR = \frac{1}{N} \sum_{i=1}^{N} 
\frac{1}{\text{rank}_i}
\label{mrr}
\end{equation}
where $N$ denotes the total number of bug reports. $\text{rank}_i$ refers to the rank of the first correct buggy file for the i-th bug report. The higher the value of MRR, the better the technique.

% \end{enumerate}
We selected these two metrics to capture both ranking quality and practical utility. MRR emphasizes the importance of returning the first relevant result as early as possible, which is crucial for minimizing developer effort \cite{niu2025deep}. Although multiple files may be associated with a single bug report, identifying the first correct file is especially important, as it often serves as a starting point for discovering additional faulty files through manual inspection or complementary debugging techniques \cite{pradel2020scaffle}. On the other hand, Accuracy@k is highly practical for real-world scenarios where developers typically examine the top few candidates rather than inspecting an entire ranked list \cite{niu2025deep}.

% \vspace{-5pt}
\subsection{Statistical Test}\label{3_3} To determine whether the observed performance differences are statistically significant, we conducted statistical analyses using the reciprocal rank (RR) values of individual issue reports. We first examined the normality of the RR distributions using the Anderson–Darling test, which is considered one of the most powerful tests for assessing normality \cite{razali2011power}. Since the distributions were not normally distributed, we employed non-parametric statistical tests. The choice of statistical test depended on the relationship between the observations being compared. For all analyses, we used a significance level of 0.05 \cite{wasserstein2016asa}.

\textbf{Intra-technique comparison:} We conducted two types of intra-technique comparisons to examine the impact of mislabeled issues in the test set and historical/training data separately.

(i) ORIG vs. TC: To assess the impact of mislabeled test issues, we compared each technique's performance between the Original (ORIG) and Test-Cleaned (TC) configurations. Since TC is a subset of ORIG, the two samples are partially overlapping: genuine bug reports appear in both configurations and form paired observations, whereas non-bug reports appear only in ORIG and form unpaired observations. Consequently, conventional non-parametric tests are not directly applicable. The Wilcoxon signed-rank test requires fully paired observations \cite{rosner2006wilcoxon}, while the Mann–Whitney U test assumes fully independent samples \cite{nachar2008mann}.

To address this, we applied the non-parametric partially overlapping samples test statistic, as proposed in \cite{derrick2020parametric}. There are two variants of this test statistic: one assumes ($T_{\text{RNK1}}$) equal variances and another ($T_{\text{RNK2}}$) does not. To determine the appropriate choice, we conducted the Brown and Forsythe test, a robust version of Levene’s test to assess the equality of variances between the noisy and clean datasets \cite{02462954-2341-35c6-ac32-7bc139c0d8ab}. Based on the results of the Brown–Forsythe test, we applied $T_{\text{RNK1}}$ when the equal-variance assumption was satisfied and $T_{\text{RNK2}}$ otherwise. 

These tests first pool all observations from both samples into a single dataset and rank them in ascending order using the RT-1 procedure \cite{conover1981rank}. In cases of tied ranks, the median rank is assigned. Next, these rank values are used to calculate the means ($\bar{X}_1$ and $\bar{X}_2$) and variances ($S_1^2$ and $S_2^2$) for the ORIG and TC data. Finally, these values are used in equation \ref{eq:trnk1} and \ref{eq:trnk2} for calculating $T_{\text{RNK1}}$ and $T_{\text{RNK2}}$ respectively.

\begin{align}
T_{\text{RNK1}} &= 
\frac{\bar{X}_1 - \bar{X}_2}{
S_p \sqrt{
\frac{1}{n_1} + \frac{1}{n_2} 
- 2r \left( \frac{n_c}{n_1 n_2} \right)
}
} \label{eq:trnk1} \\
S_p &= \sqrt{
\frac{(n_1 - 1)S_1^2 + (n_2 - 1)S_2^2}{
n_1 + n_2 - 2
}
}
\end{align}

\begin{align}
T_{\text{RNK2}} &= 
\frac{\bar{X}_1 - \bar{X}_2}{
\sqrt{
\frac{S_1^2}{n_1} + \frac{S_2^2}{n_2} 
- 2r \left( \frac{S_1 S_2 n_c}{n_1 n_2} \right)
}
} \label{eq:trnk2}
\end{align}

% \begin{equation}
% T_{\text{RNK2}}
% =
% \tfrac{\bar{X}_1-\bar{X}_2}
% {\sqrt{\tfrac{S_1^2}{n_1}+\tfrac{S_2^2}{n_2}-2r\left(\tfrac{S_1S_2n_c}{n_1n_2}\right)}}.
% \label{eq:trnk2}
% \end{equation}

The test statistic ($T_{\text{RNK1}}$ or $T_{\text{RNK2}}$) is referenced against the t-distribution with degrees of freedom $\nu_2$, calculated using equation \ref{eq:v2}.

\begin{equation}
\label{eq:v2}
\nu_2 = (n_c - 1) + 
\left( \frac{\gamma - n_c + 1}{n_a + n_b + 2n_c} \right)(n_a + n_b)
\end{equation}

\begin{equation}
\label{eq:gamma_compact}
\gamma = 
\frac{\left( S_1^2 / n_1 + S_2^2 / n_2 \right)^2}
     {\left( S_1^2 / n_1 \right)^2 / (n_1 - 1) + \left( S_2^2 / n_2 \right)^2 / (n_2 - 1)}
\end{equation}

\vspace{0.5em}
In the equations,

\noindent
\begin{tabular}{@{}>{\raggedleft\arraybackslash}p{0.8cm} p{7cm}@{}}   % Adjust widths as needed
% \toprule
% \textbf{Symbol} & \textbf{Meaning} \\
% \midrule
$\bar{X}_1,\, \bar{X}_2$ & Means of ORIG and TC data respectively \\
$n_a,\, n_b$             & Number of unpaired observations exclusive to ORIG and TC data respectively. In our case, $n_b$ is 0 as there is no issue exclusive to TC data\\
$n_c$                   & Number of paired observations (bug issues) \\
$n_1$       & Total number of observations in ORIG data $(n_1 = n_a + n_c)$\\
$n_2$       & Total number of observations in TC data $(n_2 = n_b + n_c)$\\
$S_1^2,\, S_2^2$         & Sample variances of ORIG data and TC data (with Bessel's correction) \\
% $\nu_1,\, \nu_2$         & Degrees of freedom corresponding to $T_{\text{RNK1}}$ and $T_{\text{RNK2}}$ \\
% $S_p$                   & Pooled standard deviation assuming equal variances \\
$r$                     & Spearman’s rank correlation coefficient for the $n_c$ paired observations \\

% \bottomrule
\end{tabular}
\vspace{0.5em}

For effect size, we calculated Cohen’s d using equation \ref{eq:cohend}, as suggested by \cite{derrick2020statistical}.

\vspace{-3pt}

\begin{equation}
\label{eq:cohend}
d = 2t / \sqrt{\nu}
\end{equation}

% \begin{align}
% d = 2t / \sqrt{\nu} \label{eq:cohend}
% \end{align}

where $t$ denotes the test-statistic ($T_{\text{RNK1}}$ or $T_{\text{RNK2}}$) and $\nu$ is the degrees of freedom ($\nu_2$).
We interpreted the effect size using thresholds from prior work \cite{kochhar2014potential}: $d < 0.2$ (trivial); $0.2 \leq d < 0.5$ (small); $0.5 \leq d < 0.8$ (medium); $0.8 \leq d < 1.3$ (large); and $d \geq 1.3$ (very large).

% \begin{itemize}
%     \item $d < 0.2$ means trivial, 
%     \item $0.2 \leq d < 0.5$ means small, 
%     \item $0.5 \leq d < 0.8$ means medium,
%     \item $0.80 \leq d < 1.3$ means large and
%     \item $d > 1.3$ means very large. 
% \end{itemize}

To assess the robustness of the ORIG–TC findings to random classification errors in GenTag's predictions, we complemented the statistical tests above with a Monte Carlo simulation \cite{raychaudhuri2008introduction}. GenTag may occasionally misclassify issues by labeling bug reports as non-bugs or non-bug reports as bugs. We simulated both types of errors by randomly flipping issue labels. Specifically, we varied the false-negative and false-positive rates independently over five levels (1\%, 3\%, 5\%, 8\% and 10\%), resulting in 25 (5×5) error-rate combinations. The upper bound of 10\% was selected because it exceeds GenTag's observed classification error (weighted F1-score and accuracy of 0.93 on our manually validated sample). For each scenario, we generated 1,000 perturbed datasets, recomputed the MRR changes and repeated the partially overlapping samples test, resulting in 25,000 perturbed datasets and 150,000 technique-level evaluations.

(ii) TC vs. FC: To assess the impact of mislabeled historical/training issues, we compared the Test-Cleaned (TC) and Fully-Cleaned (FC) configurations for BugLocator, BRTracer, and DreamLoc. Unlike ORIG–TC, TC and FC contain exactly the same evaluation reports, with differences arising only from the historical/training data used by the techniques. Therefore, their per-issue RR values form fully paired samples. We applied the Wilcoxon signed-rank test [52] to evaluate whether the performance differences were statistically significant. We additionally quantified the magnitude of these differences using rank-biserial correlation \cite{kerby2014simple}, as defined in equation \ref{eq:rbiserial}.

\begin{equation}
\label{eq:rbiserial}
\begin{split}
rb &= \frac{W_+ - W_-}{W_+ + W_-}
\end{split}
\end{equation}

where $W_+$ and $W_-$ denote the sums of the positive and negative signed ranks respectively. 
The rank-biserial correlation $rb$ ranges from $-1$ to $1$. We interpret effect sizes following thresholds from prior literature \cite{manchala2022diversity}: $|rb| < 0.1$ (trivial); $0.1 \leq |rb| < 0.3$ (small); $0.3 \leq |rb| < 0.5$ (medium); $|rb| \geq 0.5$ (large).

\textbf{Inter-technique comparison:} To investigate whether mislabeled test issues affect the relative performance of IRBL techniques, we performed pairwise comparisons among the six techniques separately on ORIG and TC. Within each configuration, every pair of techniques was evaluated on the same set of issue reports, producing paired RR observations. Therefore, we employed the Wilcoxon signed-rank test to determine whether the differences between techniques were statistically significant. To control the family-wise error rate arising from multiple pairwise comparisons, we applied the Holm–Bonferroni correction \cite{abdi2010holm} separately to the 15 comparisons within each configuration, using a significance level of 0.05. We additionally computed rank-biserial correlation, as defined in equation \ref{eq:rbiserial}, to quantify the magnitude and direction of the differences. By comparing the statistical significance and effect sizes obtained on ORIG and TC, we assessed whether removing mislabeled issues changes the relative performance differences among IRBL techniques and the magnitude of such changes.

\begin{table*}[]
\centering
\caption{Intra Technique Comparison using Accuracy@k across ORIG \& TC}
\tabcolsep=0.25cm
\begin{tabular}{|l|rrr|rrr|rrr|}
\hline
\multirow{2}{*}{\textbf{Technique}} & \multicolumn{3}{c|}{\textbf{Accuracy@1}}                                           & \multicolumn{3}{c|}{\textbf{Accuracy@5}}                                           & \multicolumn{3}{c|}{\textbf{Accuracy@10}}                                          \\ \cline{2-10} 
                                    & \multicolumn{1}{c}{ORIG} & \multicolumn{1}{c}{TC} & \multicolumn{1}{c|}{Diff (\%)} & \multicolumn{1}{c}{ORIG} & \multicolumn{1}{c}{TC} & \multicolumn{1}{c|}{Diff (\%)} & \multicolumn{1}{c}{ORIG} & \multicolumn{1}{c}{TC} & \multicolumn{1}{c|}{Diff (\%)} \\ \hline
VSM                                 & 3.52\%                   & 2.49\%                 & -29.26\%                       & 8.53\%                   & 6.62\%                 & -22.39\%                       & 12.17\%                  & 9.62\%                 & -20.95\%                       \\ \hline
BugLocator                          & 23.37\%                  & 20.96\%                & -10.31\%                       & 45.32\%                  & 42.11\%                & -7.08\%                        & 55.57\%                  & 52.57\%                & -5.40\%                        \\ \hline
BLUiR                               & 25.33\%                  & 22.95\%                & -9.40\%                        & 48.19\%                  & 45.24\%                & -6.12\%                        & 58.11\%                  & 54.89\%                & -5.54\%                        \\ \hline
BRTracer                            & 24.95\%                  & 22.45\%                & -10.02\%                       & 47.96\%                  & 44.81\%                & -6.57\%                        & 58.07\%                  & 54.98\%                & -5.32\%                        \\ \hline
DreamLoc                            & 28.62\%                  & 26.77\%                & -6.46\%                        & 54.52\%                  & 52.35\%                & -3.98\%                        & 64.74\%                  & 63.23\%                & -2.33\%                        \\ \hline
GenLoc                              & 40.55\%                  & 38.36\%                & -5.40\%                        & 61.24\%                  & 58.49\%                & -4.49\%                        & 67.00\%                  & 64.41\%                & -3.87\%                        \\ \hline
\end{tabular}
\label{tab:intra-accuracy}
\end{table*}

\subsection{Result Analysis of IRBL Approaches}

\textit{Intra-technique Performance:} We present the intra-technique performance of the evaluated IRBL approaches under the three dataset configurations: Original (ORIG), Test-Cleaned (TC), and Fully-Cleaned (FC). ORIG corresponds to the benchmark as originally released, TC removes non-bug reports only from the test partition, and FC removes non-bug reports from both the historical/training and test partitions. This design enables us to distinguish the effect of mislabeled issues in the evaluation set from their effect in historical or training data.

We first compare ORIG and TC to examine how mislabeled issues in the test set affect the measured performance of IRBL techniques. Tables~\ref{tab:intra-accuracy} and~\ref{tab:intra-mrr} present the corresponding Accuracy@k and MRR results, respectively. Across all six approaches, removing non-bug reports from the test set consistently reduces measured performance across all metrics. The reduction is most pronounced for VSM, whose Accuracy@1 decreases by 29.26\%. The remaining techniques also exhibit reductions in Accuracy@1: 10.31\% for BugLocator, 9.40\% for BLUiR, 10.02\% for BRTracer, 6.46\% for DreamLoc, and 5.40\% for GenLoc. These declines are notably larger than those reported in prior studies~\cite{kochhar2014s,widyasari2022influence}, which observed only minimal reductions in Accuracy@1 of classical techniques after removing mislabeled issues. This difference suggests that the impact of mislabeled issues may be more pronounced when IRBL techniques are evaluated on a large-scale and diverse benchmark such as the one considered in our study. The reductions at early cutoff values are also practically relevant, as 83.01\% of practitioners consider a bug successfully localized when the correct file appears within the top five ranked results~\cite{kochhar2016practitioners}. Therefore, even moderate reductions in early-ranked localization performance can affect the practical usefulness of IRBL techniques.

A similar pattern emerges for MRR. The differences in reciprocal rank between ORIG and TC are statistically significant for all six techniques, with effect sizes ranging from 0.15 to 0.26. Unlike prior studies~\cite{kochhar2014s,widyasari2022influence}, which reported statistically insignificant differences for VSM, we observe a statistically significant and substantial reduction (21.74\%) in its MRR. BugLocator, BLUiR, BRTracer, DreamLoc, and GenLoc likewise exhibit statistically significant reductions. Notably, even the recent approaches evaluated in our study, including the deep learning-based DreamLoc and LLM-based GenLoc, are affected by mislabeled test issues.

\begin{figure}[htbp]
\centering
\begin{tikzpicture}[scale=0.67]
\begin{axis}[
    xlabel={rank},
    ylabel={reciprocal rank},
    domain=1:100,
    samples=500,
    ymin=0, ymax=1.05,
    xmin=0, xmax=105,
    width=\columnwidth,
    height=0.6\columnwidth,
    grid=both,
    major grid style={line width=0.4pt,draw=gray!50},
    minor grid style={line width=0.2pt,draw=gray!20},
    thin,
    smooth,
    axis lines=left,
    tick align=outside,
    tick style={black},
    xtick distance=20,
    ytick distance=0.25,
]

\addplot[
    blue,
    thin,
    smooth
]
{1/x};

\end{axis}
\end{tikzpicture}
\caption{Reciprocal Rank Decreases with Increasing Rank}
\label{fig:rr-curve}
\end{figure}
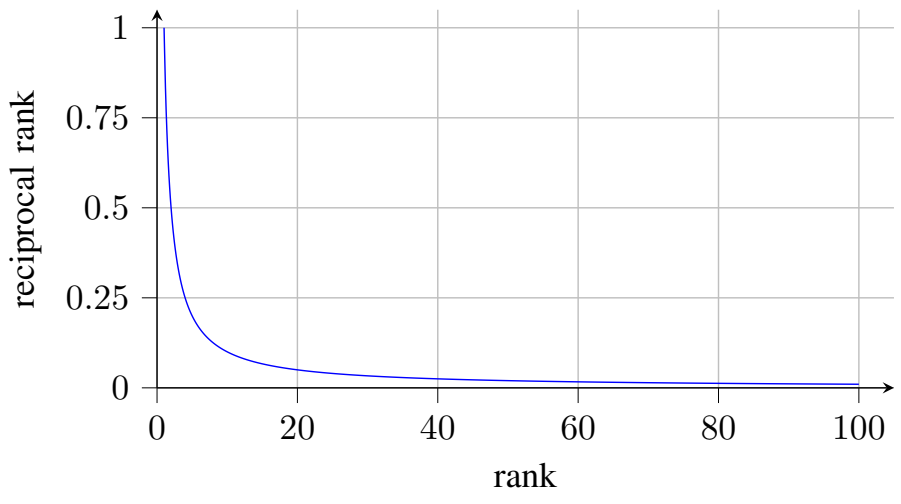

To better understand why VSM exhibits the largest relative decline in MRR, we first consider the nature of the MRR metric. As illustrated in Fig.~\ref{fig:rr-curve}, reciprocal rank decreases sharply at early positions. For example, moving the first correct file from rank 1 to rank 5 causes a substantially larger reduction in reciprocal rank than moving it from rank 40 to rank 60. Consequently, a technique that loses a considerable number of early-ranked successful predictions after test-set cleaning can experience a pronounced decline in MRR.

\begin{table}[]
\centering
\caption{Intra Technique Comparison using MRR Across ORIG \& TC}
\tabcolsep=0.12cm
\begin{tabular}{|l|r|r|r|c|c|}
\hline
\multicolumn{1}{|c|}{\textbf{Technique}} & \multicolumn{1}{c|}{\textbf{\begin{tabular}[c]{@{}c@{}}ORIG\\ MRR\end{tabular}}} & \multicolumn{1}{c|}{\textbf{\begin{tabular}[c]{@{}c@{}}TC\\ MRR\end{tabular}}} & \multicolumn{1}{c|}{\textbf{Diff (\%)}} & \textbf{p-value} & \textbf{Effect size} \\ \hline
VSM                                      & 0.069                                                                                & 0.054                                                                                    & -21.74\%                                & <0.001             & 0.26                 \\ \hline
BugLocator                               & 0.341                                                                                & 0.314                                                                                    & -7.92\%                                 & <0.001             & 0.22                 \\ \hline
BLUiR                                    & 0.362                                                                                & 0.335                                                                                    & -7.46\%                                 & <0.001             & 0.23                 \\ \hline
BRTracer                                 & 0.360                                                                                & 0.333                                                                                    & -7.50\%                                 & <0.001             & 0.22                 \\ \hline
DreamLoc                                 & 0.406                                                                                & 0.387                                                                                    & -4.68\%                                 & <0.001             & 0.15                 \\ \hline
GenLoc                                   & 0.493                                                                                & 0.469                                                                                    & -4.87\%                                 & <0.001             & 0.17                 \\ \hline
\end{tabular}
\label{tab:intra-mrr}
\end{table}

We therefore examine the contribution of bug and non-bug reports to successful predictions at ranks 1--5. Table \ref{top-rank} shows that VSM disproportionately benefits from non-bug reports. Although non-bug reports constitute only about one-third of the benchmark, they account for 53\% of VSM's correctly localized top-1 cases, 51\% at rank 2, and approximately half of its successful predictions across ranks 3--5. In contrast, genuine bug reports contribute the majority of early-ranked successful predictions for all other techniques. At rank 1, bug reports account for 59\% of BugLocator's, 60\% of BLUiR's and BRTracer's, 62\% of DreamLoc's, and 63\% of GenLoc's successful predictions. This disproportionate contribution of non-bug reports helps explain why VSM experiences a substantially larger performance reduction when such reports are excluded from the test set.

\begin{table}[htb]
\centering
\caption{Contribution of Bug/Non-Bug Issues to Top Ranks}
\tabcolsep=0.22cm
\begin{tabular}{|l|l|rrrrr|}
\hline
\multicolumn{1}{|c|}{\multirow{2}{*}{\textbf{Technique}}} & \multicolumn{1}{c|}{\multirow{2}{*}{\textbf{Type}}} & \multicolumn{5}{c|}{\textbf{Rank}}                                                                                                                                           \\ \cline{3-7} 
\multicolumn{1}{|c|}{}                                    & \multicolumn{1}{c|}{}                              & \multicolumn{1}{c|}{\textbf{@1}} & \multicolumn{1}{c|}{\textbf{@2}} & \multicolumn{1}{c|}{\textbf{@3}} & \multicolumn{1}{c|}{\textbf{@4}} & \multicolumn{1}{c|}{\textbf{@5}} \\ \hline
\multirow{2}{*}{VSM}                                      & bug                                               & \multicolumn{1}{r|}{47}        & \multicolumn{1}{r|}{49}        & \multicolumn{1}{r|}{51}        & \multicolumn{1}{r|}{50}        & 51                             \\ \cline{2-7} 
                                                          & non-bug                                           & \multicolumn{1}{r|}{53}        & \multicolumn{1}{r|}{51}        & \multicolumn{1}{r|}{49}        & \multicolumn{1}{r|}{50}        & 49                             \\ \hline
\multirow{2}{*}{BugLocator}                               & bug                                               & \multicolumn{1}{r|}{59}        & \multicolumn{1}{r|}{60}        & \multicolumn{1}{r|}{61}        & \multicolumn{1}{r|}{61}        & 62                             \\ \cline{2-7} 
                                                          & non-bug                                           & \multicolumn{1}{r|}{41}        & \multicolumn{1}{r|}{40}        & \multicolumn{1}{r|}{39}        & \multicolumn{1}{r|}{39}        & 38                             \\ \hline
\multirow{2}{*}{BLUiR}                                    & bug                                               & \multicolumn{1}{r|}{60}        & \multicolumn{1}{r|}{61}        & \multicolumn{1}{r|}{61}        & \multicolumn{1}{r|}{62}        & 62                             \\ \cline{2-7} 
                                                          & non-bug                                           & \multicolumn{1}{r|}{40}        & \multicolumn{1}{r|}{39}        & \multicolumn{1}{r|}{39}        & \multicolumn{1}{r|}{38}        & 38                             \\ \hline
\multirow{2}{*}{BRTracer}                                 & bug                                               & \multicolumn{1}{r|}{60}        & \multicolumn{1}{r|}{60}        & \multicolumn{1}{r|}{61}        & \multicolumn{1}{r|}{62}        & 62                             \\ \cline{2-7} 
                                                          & non-bug                                           & \multicolumn{1}{r|}{40}        & \multicolumn{1}{r|}{40}        & \multicolumn{1}{r|}{39}        & \multicolumn{1}{r|}{38}        & 38                             \\ \hline
\multirow{2}{*}{DreamLoc}                                 & bug                                               & \multicolumn{1}{r|}{62}        & \multicolumn{1}{r|}{63}        & \multicolumn{1}{r|}{63}        & \multicolumn{1}{r|}{63}        & 64                             \\ \cline{2-7} 
                                                          & non-bug                                           & \multicolumn{1}{r|}{38}        & \multicolumn{1}{r|}{37}        & \multicolumn{1}{r|}{37}        & \multicolumn{1}{r|}{37}        & 36                             \\ \hline
\multirow{2}{*}{GenLoc}                                 & bug                                               & \multicolumn{1}{r|}{63}        & \multicolumn{1}{r|}{63}        & \multicolumn{1}{r|}{63}        & \multicolumn{1}{r|}{63}        & 63                             \\ \cline{2-7} 
                                                          & non-bug                                           & \multicolumn{1}{r|}{37}        & \multicolumn{1}{r|}{37}        & \multicolumn{1}{r|}{37}        & \multicolumn{1}{r|}{37}        & 37                             \\ \hline
\end{tabular}
\label{top-rank}
\end{table}

Manual examination of representative cases provides further insight into this pattern. We observed that non-bug reports often contain terms such as UI labels, method names, or field identifiers that closely match source-code tokens, making them comparatively easy for lexical retrieval techniques such as VSM to localize. Fig.~\ref{birt_issue} illustrates such a non-bug issue that requests a UI label change from "Data Field" to a more accurate alternative. Because the report directly mentions a term appearing in the relevant source code, VSM assigns a high rank to the associated file. When such non-bug reports are removed from the evaluation set, VSM loses a substantial portion of its early-ranked successful predictions, contributing to its larger decline in both Accuracy@k and MRR.

\begin{figure}[htb]
\centering
\begin{tcolorbox}[
  top=2pt,
  bottom=2pt,
  left=2pt,
  right=2pt,
  % boxsep=4pt,
  colback=gray!10!white, 
  colframe=black!75!black, 
]
\small
\raggedright
\fontsize{7.8pt}{7.8pt}\selectfont
\textbf{Summary:} BIRT aggregate dialog "Data Field" label incorrect\\
\textbf{Description:} Open or create a BIRT report. Add a table. Open the Bindings tab in the table property editor. Click on the "Add Aggregate" button to add an aggregate. Notice that the fifth field is "Data Field:" This is a bit of a misnomer. Using the expression builder, I can reference a data field. I can call a JavaScript function. I can reference a column binding. Or, I can create an expression that includes all of the above. "Expression:" would be a better name for this label. Or perhaps "Value:". Or even the awkward "Value Expression:".
\end{tcolorbox}
\caption{Non-bug Issue (ID: 224502) from Birt Project}
\label{birt_issue}
\end{figure}

Although the observed effect sizes are generally small, the differences in measured localization performance can still have meaningful practical implications. File-level bug localization is inherently coarse-grained, requiring developers to manually inspect candidate files to identify the faulty method, statement, or expression~\cite{wen2016locus}. When the correct buggy file is ranked lower even by a few positions, developers must examine more files, each of which demands time and cognitive effort to understand. Consequently, even moderate rank degradations translate into greater debugging effort. Moreover, degradation in bug localization can affect downstream tasks such as Automated Program Repair (APR)~\cite{le2019automated,hossain2024deep}. When the rank of the correct buggy file degrades, APR tools may generate invalid patches by modifying irrelevant parts of the codebase, thereby reducing their effectiveness \cite{assiri2017fault,liu2019you}. For example, an APR tool that relies on GenLoc and is calibrated based on its performance on the original noisy dataset may fail in practice if that performance does not hold on cleaner data. In such cases, the correct file may fall outside the top-ranked candidates, causing the APR tool to apply transformations to unrelated files and produce invalid patches. Therefore, reliable evaluation of bug localization techniques is important not only for fair benchmarking but also for downstream software maintenance tasks that depend on localization quality.

To assess whether the observed ORIG--TC differences are robust to random classification errors in GenTag's predictions, we conducted the Monte Carlo analysis described in Section~\ref{3_3}. Across all 25,000 simulated datasets, the performance decrease remained consistent, and statistical significance (\(p<0.05\)) was maintained in 100\% of trials for every technique across all considered error-rate combinations. The 95\% confidence intervals for the relative MRR change were narrow with both the lower and upper bounds remaining negative for every technique: VSM [-22.17\%, -16.52\%], BugLocator [-8.31\%, -6.10\%], BLUiR [-7.70\%, -5.60\%], BRTracer [-7.92\%, -5.79\%], DreamLoc [-5.18\%, -3.45\%], and GenLoc [-5.23\%, -3.64\%]. These results indicate that the observed reduction in measured performance after test-set cleaning is robust to plausible GenTag classification errors.

\begin{table*}[]
\centering
\caption{Intra Technique Comparison using Accuracy@k Across TC \& FC}
\tabcolsep=0.25cm
\begin{tabular}{|l|rrr|rrr|rrr|}
\hline
\multirow{2}{*}{\textbf{Technique}} & \multicolumn{3}{c|}{\textbf{Accuracy@1}}                                           & \multicolumn{3}{c|}{\textbf{Accuracy@5}}                                           & \multicolumn{3}{c|}{\textbf{Accuracy@10}}                                          \\ \cline{2-10} 
                                    & \multicolumn{1}{c}{TC} & \multicolumn{1}{c}{FC} & \multicolumn{1}{c|}{Diff (\%)} & \multicolumn{1}{c}{TC} & \multicolumn{1}{c}{FC} & \multicolumn{1}{c|}{Diff (\%)} & \multicolumn{1}{c}{TC} & \multicolumn{1}{c}{FC} & \multicolumn{1}{c|}{Diff (\%)} \\ \hline
BugLocator                          & 20.96\%                  & 20.81\%                & -0.72\%                        & 42.11\%                  & 41.58\%                & -1.26\%                        & 52.57\%                  & 52.09\%                & -0.91\%                        \\ \hline
BRTracer                            & 22.45\%                  & 22.82\%                & 1.65\%                         & 44.81\%                  & 44.81\%                & 0.00\%                         & 54.98\%                  & 55.21\%                & 0.42\%                         \\ \hline
DreamLoc                            & 26.77\%                  & 23.53\%                & -12.10\%                       & 52.35\%                  & 47.40\%                & -9.46\%                        & 63.23\%                  & 58.49\%                & -7.50\%                        \\ \hline
\end{tabular}
\label{tab:intra-accuracy-tc-fc}
\end{table*}

We next compare TC and FC to examine the effect of mislabeled issues in historical or training data. Tables~\ref{tab:intra-accuracy-tc-fc} and~\ref{tab:intra-mrr-tc-fc} present the corresponding Accuracy@k and MRR results, respectively. This comparison applies only to techniques that rely on such information, namely BugLocator, BRTracer, and DreamLoc. Unlike the consistent reductions observed when cleaning the test set, the effect of cleaning historical/training data varies considerably across techniques. BugLocator exhibits minor decreases across all metrics, whereas BRTracer shows slight improvements or unchanged performance, indicating that both techniques are relatively insensitive to the removal of mislabeled historical reports. In contrast, DreamLoc experiences a more pronounced decline, with Accuracy@1 decreasing by 12.10\% and MRR by 6.98\%. One possible explanation is that DreamLoc, as a deep learning-based approach, benefits from a larger volume of training data, even when some reports are mislabeled. Removing these reports reduces the available training data, potentially limiting the model's ability to learn useful relevance patterns. These results show that, unlike test-set noise, whose removal consistently lowers measured performance across all evaluated techniques, the effect of mislabeled issues in historical or training data is technique-dependent.

Since TC and FC are evaluated on exactly the same set of genuine bug reports, we apply the Wilcoxon signed-rank test to the paired reciprocal-rank observations. The results indicate statistically significant differences for the three history-dependent techniques. However, the magnitude and direction of the changes differ across approaches, reinforcing that mislabeled historical/training data does not have a uniform effect on IRBL techniques. The underlying reasons for these differences may depend on how individual approaches exploit historical or training information and are beyond the scope of this study.

Overall, our results reveal two distinct effects of mislabeled issues. First, mislabeled reports in the test set systematically inflate the measured performance of all evaluated IRBL techniques, although the magnitude of this inflation varies across approaches. Second, mislabeled reports in historical or training data have a technique-dependent effect: BugLocator and BRTracer remain largely stable after cleaning, whereas DreamLoc exhibits a more pronounced reduction. These findings demonstrate the importance of distinguishing between test-set contamination and historical/training-set contamination when assessing the impact of mislabeled issues on IRBL evaluations.

% \vspace{-4pt}
\begin{table}[h]
\centering
\caption{Intra Technique Comparison using MRR Across TC \& FC}
\tabcolsep=0.12cm
\begin{tabular}{|l|r|r|r|c|c|}
\hline
\multicolumn{1}{|c|}{\textbf{Technique}} & \multicolumn{1}{c|}{\textbf{\begin{tabular}[c]{@{}c@{}}TC\\ MRR\end{tabular}}} & \multicolumn{1}{c|}{\textbf{\begin{tabular}[c]{@{}c@{}}FC\\ MRR\end{tabular}}} & \multicolumn{1}{c|}{\textbf{Diff (\%)}} & \textbf{p-value} & \textbf{Effect size} \\ \hline
BugLocator                               & 0.314                                                                          & 0.311                                                                          & -0.96\%                                 & <0.001             & 0.141                \\ \hline
BRTracer                                 & 0.333                                                                          & 0.334                                                                          & 0.30\%                                  & <0.001             & 0.216                \\ \hline
DreamLoc                                 & 0.387                                                                          & 0.360                                                                           & -6.98\%                                 & <0.001             & -0.121               \\ \hline
\end{tabular}
\label{tab:intra-mrr-tc-fc}
\end{table}

\begin{table*}[t]
\centering
\caption{Inter Technique Comparison using MRR}
\small
\begin{tabular}{|l|l|rcrc|rcrc|}
\hline
\multicolumn{1}{|c|}{\multirow{2}{*}{\textbf{Technique 1}}} & \multicolumn{1}{c|}{\multirow{2}{*}{\textbf{Technique 2}}} & \multicolumn{4}{c|}{\textbf{Original}}                                                                                    & \multicolumn{4}{c|}{\textbf{Test-Cleaned}}                                                                                   \\ \cline{3-10} 
\multicolumn{1}{|c|}{}                                      & \multicolumn{1}{c|}{}                                      & \multicolumn{1}{c|}{p-value} & \multicolumn{1}{c|}{Significant} & \multicolumn{1}{c|}{Effect Size}   & Interpretation  & \multicolumn{1}{c|}{p-value} & \multicolumn{1}{c|}{Significant} & \multicolumn{1}{c|}{Effect Size}   & Interpretation \\ \hline
VSM                                                         & BugLocator                                                 & \multicolumn{1}{r|}{<0.001}     & \multicolumn{1}{c|}{Yes}         & \multicolumn{1}{r|}{0.87}          & large           & \multicolumn{1}{r|}{<0.001}     & \multicolumn{1}{c|}{Yes}         & \multicolumn{1}{r|}{0.88}          & large          \\ \hline
VSM                                                         & BLUiR                                                      & \multicolumn{1}{r|}{<0.001}     & \multicolumn{1}{c|}{Yes}         & \multicolumn{1}{r|}{0.86}          & large           & \multicolumn{1}{r|}{<0.001}     & \multicolumn{1}{c|}{Yes}         & \multicolumn{1}{r|}{0.88}          & large          \\ \hline
VSM                                                         & BRTracer                                                   & \multicolumn{1}{r|}{<0.001}     & \multicolumn{1}{c|}{Yes}         & \multicolumn{1}{r|}{0.87}          & large           & \multicolumn{1}{r|}{<0.001}     & \multicolumn{1}{c|}{Yes}         & \multicolumn{1}{r|}{0.89}          & large          \\ \hline
VSM                                                         & DreamLoc                                                   & \multicolumn{1}{r|}{<0.001}     & \multicolumn{1}{c|}{Yes}         & \multicolumn{1}{r|}{0.89}          & large           & \multicolumn{1}{r|}{<0.001}     & \multicolumn{1}{c|}{Yes}         & \multicolumn{1}{r|}{0.91}          & large          \\ \hline
VSM                                                         & GenLoc                                                     & \multicolumn{1}{r|}{<0.001}     & \multicolumn{1}{c|}{Yes}         & \multicolumn{1}{r|}{0.74}          & large           & \multicolumn{1}{r|}{<0.001}     & \multicolumn{1}{c|}{Yes}         & \multicolumn{1}{r|}{0.71}          & large          \\ \hline
BugLocator                                                  & BLUiR                                                      & \multicolumn{1}{r|}{<0.001}     & \multicolumn{1}{c|}{Yes}         & \multicolumn{1}{r|}{0.09}          & trivial         & \multicolumn{1}{r|}{<0.001}     & \multicolumn{1}{c|}{Yes}         & \multicolumn{1}{r|}{0.09}          & trivial        \\ \hline
BugLocator                                                  & BRTracer                                                   & \multicolumn{1}{r|}{<0.001}     & \multicolumn{1}{c|}{Yes}         & \multicolumn{1}{r|}{0.17}          & small           & \multicolumn{1}{r|}{<0.001}     & \multicolumn{1}{c|}{Yes}         & \multicolumn{1}{r|}{0.19}          & small          \\ \hline
BugLocator                                                  & DreamLoc                                                   & \multicolumn{1}{r|}{<0.001}     & \multicolumn{1}{c|}{Yes}         & \multicolumn{1}{r|}{0.22}          & small           & \multicolumn{1}{r|}{<0.001}     & \multicolumn{1}{c|}{Yes}         & \multicolumn{1}{r|}{0.26}          & small          \\ \hline
BugLocator                                                  & GenLoc                                                     & \multicolumn{1}{r|}{<0.001}     & \multicolumn{1}{c|}{Yes}         & \multicolumn{1}{r|}{0.35}          & medium          & \multicolumn{1}{r|}{<0.001}     & \multicolumn{1}{c|}{Yes}         & \multicolumn{1}{r|}{0.34}          & medium         \\ \hline
BLUiR                                                       & BRTracer                                                   & \multicolumn{1}{r|}{0.060}    & \multicolumn{1}{c|}{No}          & \multicolumn{1}{r|}{-0.03}         & trivial         & \multicolumn{1}{r|}{0.243}    & \multicolumn{1}{c|}{No}          & \multicolumn{1}{r|}{-0.02}         & trivial        \\ \hline
BLUiR                                                       & DreamLoc                                                   & \multicolumn{1}{r|}{<0.001}     & \multicolumn{1}{c|}{Yes}         & \multicolumn{1}{r|}{0.16}          & small           & \multicolumn{1}{r|}{<0.001}     & \multicolumn{1}{c|}{Yes}         & \multicolumn{1}{r|}{0.20}          & small          \\ \hline
BLUiR                                                       & GenLoc                                                     & \multicolumn{1}{r|}{<0.001}     & \multicolumn{1}{c|}{Yes}         & \multicolumn{1}{r|}{0.28}          & small           & \multicolumn{1}{r|}{<0.001}     & \multicolumn{1}{c|}{Yes}         & \multicolumn{1}{r|}{0.27}          & small          \\ \hline
BRTracer                                                    & DreamLoc                                                   & \multicolumn{1}{r|}{<0.001}     & \multicolumn{1}{c|}{Yes}         & \multicolumn{1}{r|}{0.16}          & small           & \multicolumn{1}{r|}{<0.001}     & \multicolumn{1}{c|}{Yes}         & \multicolumn{1}{r|}{0.20}          & small          \\ \hline
BRTracer                                          & GenLoc                                            & \multicolumn{1}{r|}{<0.001}     & \multicolumn{1}{c|}{Yes}         & \multicolumn{1}{r|}{0.31} & medium & \multicolumn{1}{r|}{<0.001}     & \multicolumn{1}{c|}{Yes}         & \multicolumn{1}{r|}{0.30} & small \\ \hline
DreamLoc                                                    & GenLoc                                                     & \multicolumn{1}{r|}{<0.001}     & \multicolumn{1}{c|}{Yes}         & \multicolumn{1}{r|}{0.16}          & small           & \multicolumn{1}{r|}{<0.001}     & \multicolumn{1}{c|}{Yes}         & \multicolumn{1}{r|}{0.13}          & small          \\ \hline
\end{tabular}
\label{tab:inter}
\end{table*}

\textit{Inter-technique Performance:} To investigate whether mislabeled issues influence the comparative evaluation of IRBL techniques, we perform pairwise comparisons using the Original (ORIG) and Test-Cleaned (TC) configurations. We select TC rather than Fully-Cleaned (FC) based on our intra-technique findings, which show that removing mislabeled historical/training reports has negligible effects on BugLocator and BRTracer, while reducing DreamLoc's performance. These findings suggest that retaining historical/training reports, including those classified as non-bugs, may provide useful information for certain techniques. Furthermore, comparing ORIG and TC allows us to isolate the impact of mislabeled evaluation reports without introducing technique-dependent effects from cleaning historical/training data. We therefore compare the reciprocal rank (RR) values of each technique pair on ORIG and TC. Table~\ref{tab:inter} presents the results of the Wilcoxon signed-rank tests, including Holm-adjusted p-values and rank-biserial correlation effect sizes. A positive effect size indicates that the second technique outperforms the first.

The results show that the statistical significance of pairwise differences remains consistent after removing mislabeled issues. Among the 15 technique pairs, 14 exhibit statistically significant differences (p < 0.001) on both ORIG and TC. The only exception is BLUiR versus BRTracer, for which the differences remain statistically insignificant (p = 0.060 on ORIG and p = 0.243 on TC). Although the significance outcomes remain unchanged, the effect sizes vary across dataset configurations, indicating that mislabeled issues influence the magnitude of performance differences between techniques.

The most pronounced differences are observed between VSM and the remaining approaches. On ORIG, the effect sizes range from 0.74 to 0.89, indicating large differences in favor of the other techniques. After cleaning, these effect sizes increase for four of the five comparisons: VSM versus BugLocator (0.87 to 0.88), BLUiR (0.86 to 0.88), BRTracer (0.87 to 0.89), and DreamLoc (0.89 to 0.91). This pattern suggests that removing mislabeled issues further widens the relative performance gap between VSM and these approaches. These findings are consistent with our intra-technique analysis, where VSM exhibits the greatest sensitivity to mislabeled test reports.

Among the more advanced techniques, the changes are generally smaller but reveal different patterns. The effect sizes for DreamLoc compared with BugLocator, BLUiR, and BRTracer increase from 0.22 to 0.26, 0.16 to 0.20, and 0.16 to 0.20, respectively, suggesting that DreamLoc's relative advantage becomes more pronounced on the cleaned dataset. Conversely, the effect sizes for GenLoc decrease slightly when compared with VSM (0.74 to 0.71), BugLocator (0.35 to 0.34), BLUiR (0.28 to 0.27), BRTracer (0.31 to 0.30), and DreamLoc (0.16 to 0.13), suggesting that GenLoc's relative advantage becomes slightly less pronounced after removing mislabeled issues. Nevertheless, GenLoc maintains a statistically significant advantage over all other techniques on both datasets.

Overall, these results demonstrate that mislabeled issues can influence the comparative evaluation of IRBL techniques. Although removing mislabeled issues does not change the statistical significance or direction of the pairwise comparisons in our evaluation, it alters the magnitude of the observed differences, sometimes widening and sometimes narrowing the gap between techniques. Consequently, conclusions about the extent to which one IRBL technique outperforms another may depend on the quality of the evaluation dataset, highlighting the importance of using correctly labeled bug reports for fair and reliable comparisons.

\section{Discussion}
This study highlights a critical yet underexamined issue in IRBL research: the widespread inclusion of mislabeled issues in benchmark datasets and the resulting implications for the reliability and practical utility of IRBL techniques.

Our analysis of five IRBL benchmarks spanning more than a decade indicates that issue mislabeling remains a persistent problem, even in recently introduced datasets. This raises concerns about the common practice of treating issue-tracker labels as reliable ground truth. In particular, including non-bug reports in evaluation datasets may lead to overly optimistic estimates of bug localization effectiveness, as our results suggest that these reports tend to be easier to localize than genuine bug reports. These findings highlight the importance of validating issue labels during benchmark construction to ensure the reliability of IRBL performance evaluations.

Importantly, our findings diverge from earlier studies \cite{kochhar2014s,widyasari2022influence} that reported only minor or statistically insignificant changes in performance (e.g., Accuracy@1, MRR), whereas our results show consistent and statistically significant performance degradation across all techniques when evaluated on clean data. This discrepancy is likely due to our use of a larger benchmark dataset, suggesting that the impact of mislabeled issues becomes more pronounced in realistic, large-scale evaluation settings.

A key insight from our study is that the impact of mislabeling depends on both the localization technique and the role of the affected data. Techniques relying primarily on lexical similarity, such as VSM, appear particularly sensitive to mislabeled evaluation reports because non-bug issues may contain explicit textual references to relevant source-code elements. Although techniques incorporating structural information, historical evidence, or learning-based representations exhibit smaller performance changes, they are not immune to evaluation noise. The impact of mislabeling also depends on whether it occurs in evaluation or historical/training data. While removing mislabeled reports from evaluation data provides a more representative assessment of localization performance on genuine bugs, removing them from historical/training data does not consistently improve performance and may even reduce it. This suggests that some reports, despite being unsuitable for evaluating bug localization, may still provide useful information for learning or historical matching. Therefore, dataset cleaning should distinguish between ensuring valid evaluation ground truth and preserving potentially useful historical/training information.

Our inter-technique analysis further demonstrates that mislabeled issues can influence the magnitude of performance differences between approaches, even when their overall ranking and statistical significance remain unchanged. This distinction is important because conclusions about the superiority of a technique often depend not only on whether it outperforms a baseline but also on the extent of that improvement. Thus, evaluations using mislabeled reports may exaggerate or underestimate the relative advantages of different IRBL techniques. These findings underscore the importance of using correctly labeled evaluation datasets and accounting for statistical dependencies when comparing localization approaches.

Finally, mislabeled issues have practical implications beyond evaluation. Performance degradation in bug localization can translate into lower-quality rankings, which prior work has shown to increase developers' code inspection effort \cite{DilshenerWY18} and negatively affect downstream tasks such as automated program repair \cite{motwani2023better}. Overall, our findings emphasize that dataset quality is a fundamental factor in IRBL research and must be carefully considered to ensure fair and meaningful evaluation.

% \clearpage

\section{Threats to Validity}
This section presents potential ways in which the validity of the study may be compromised.

\begin{description}[leftmargin=*]
    \item \textbf{Internal Validity:} A primary threat lies in the correctness of our issue classification. To address this, we evaluated our classifier against the manually labeled dataset by Herzig et al., a widely used ground truth in prior studies \cite{tantithamthavorn2015impact,kochhar2014potential,wang2018deep}. Additionally, we evaluated GenTag on a manually labeled stratified sample drawn from the five benchmark datasets and performed a Monte Carlo sensitivity analysis that showed the observed performance degradation remained statistically significant under all simulated random label perturbations. However, GenTag's classification errors may be associated with issue characteristics that also influence localization performance. Investigating the impact of such systematic errors is beyond the scope of this study and remains an avenue for future work. To reduce LLM non-determinism, we set the temperature to zero and repeated experiments three times. Although this study isolates the impact of mislabeled issues, other dataset-level biases such as tangling (i.e., incorrect ground truth) and localization hints may also influence bug localization performance. Since both the noisy and clean datasets originate from the same benchmarks, these biases are expected to remain consistent across conditions and thus do not threaten the comparative validity of our conclusions, though they may slightly affect the absolute performance values. Future work could systematically examine the joint and individual effects of these biases, e.g., by developing automated methods to detect and mitigate tangling and hint-related noise.

    \item \textbf{Construct Validity:} Construct validity concerns whether the evaluation metrics used accurately capture the performance of IRBL techniques. We mitigate this by adopting Accuracy@k and MRR, which are widely used metrics in IRBL literature \cite{kochhar2014potential, wong2023handbook}.

    \item \textbf{External Validity:} To assess the prevalence of mislabeled issues, we investigated five benchmarks spanning from 2012 to 2025. To analyze the impact of mislabeled issues, we used the widely adopted Ye et al. dataset \cite{ye2015mapping}, which contains over 22,000 reports from six diverse, large-scale Java projects. While this dataset is a standard benchmark \cite{chakraborty2024rlocator}, our findings may not generalize to industrial systems, or software written in other languages. Although we examine six representative IR-based techniques that span a range of methodologies and time frames, our results may not hold for all IRBL approaches. Further studies are required to generalize our findings to other techniques, datasets and ecosystems beyond Java.
    
\end{description}

\section{Related Work}
In the literature, there are four categories of work related to bug localization: Spectrum-Based Bug Localization (SBBL), Information Retrieval-based Bug Localization (IRBL), issue localization and issue resolution. SBBL techniques rely on execution traces from passing and failing test cases to identify suspicious program elements \cite{kang2024quantitative,rafi2024enhancing,qin2024fault,qin2025s,xu2025flexfl,shi2025enhancing}. Issue localization considers a broader range of issue types (e.g., feature requests, refactoring tasks and performance improvements) and aims to identify fine-grained code locations \cite{yu2025orcaloca,chen2025locagent,ma2508tool,chang2025bridging,jiang2025cosil}. Issue resolution focuses on generating patches to fix reported issues \cite{tao2024magis,yang2024swe,wang2024openhands,ouyang2024repograph,zhang2024autocoderover,hossain2024deep,ma2025alibaba,xia2025agentless,li2025swe}. Among all issue types, bug reports remain particularly important since unresolved bugs can compromise the reliability, performance and security of software systems \cite{kochhar2014s}. Consequently, IRBL focuses specifically on identifying buggy source files given a bug report using textual information from issue trackers. Since IRBL benchmarks assume that issues labeled as bugs correspond to actual bug reports, mislabeled issues can significantly affect the validity of IRBL evaluation. Therefore, we focus on analyzing the impact of mislabeled issues on the performance of IRBL techniques and primarily discuss prior work in IRBL.

% \paragraph{\textbf{Information Retrieval based Bug Localization}}

Existing IRBL related work can be divided into two categories: (i) novel IRBL techniques, and (ii) empirical studies analyzing them. 

Novel IRBL techniques fall into three main categories by methodology. The first category \cite{lukins2008source,nguyen2011topic,zhou2012should,sisman2012incorporating,saha2013improving,moreno2014use,wang2014version,wong2014boosting,wen2016locus,wang2016amalgam+,youm2017improved,khatiwada2018just,rath2018analyzing} uses traditional IR models such as Latent Dirichlet Allocation (LDA) and VSM. The second category  \cite{kim2013should,ye2014learning,fejzer2021tracking,lam2015combining,lam2017bug,xiao2019improving,koyuncu2019d,wang2020multi,cao2020bugpecker,qi2021dreamloc,iqbal2020determining,pradel2020scaffle,zhang2020exploiting,zakari2020multiple,anh2021imbalanced,liang2022modeling,luo2022improving,ciborowska2022fast,du2023pre,xiao2023bugradar,chakraborty2024rlocator,chakraborty2025blaze,li2025knowledge,zhou2026multi} consists of learning-based models (i.e., machine learning and deep learning models). The third category consists of reasoning-enhanced approaches \cite{Brain2025,li2025llm,caumartin2026reformulate,asad2026towards}, where LLMs are used as a reasoning component to analyze candidate files.

Several empirical studies have examined the effectiveness and limitations of IRBL techniques and datasets. Some \cite{thomas2013impact,saha2014effectiveness,akbar2020large,lee2018bench4bl} focus on evaluating techniques under different configurations of models or datasets, such as comparisons across multiple projects and languages. Other studies \cite{kochhar2014s,kochhar2014potential,widyasari2022influence,kim2021datasets,kim2022empirical,li2022empirical} highlight dataset-related issues, including mislabeled issues, incorrect ground truth and their impact on IRBL performance. In contrast, our work presents a comprehensive analysis of the impact of mislabeled issues by evaluating six representative IRBL techniques spanning the evolution of IRBL (from traditional IR models to deep learning and LLM-based approaches), conducting both intra- and inter-technique performance analyses and accounting for dataset overlap through appropriate statistical testing.

\section{Conclusion and Future Work}
This study investigates the prevalence of mislabeled issues in established benchmark datasets and their impact on IRBL techniques. Our analysis of three popular benchmark datasets, Ye et al., Bench4BL, and Zhou et al., as well as two recent datasets, GHRB and BeetleBox, shows that mislabeled issues remain widespread, accounting for 29.01\%–44.01\% of reports across datasets, with an average of 34.82\%.

To assess the impact of these mislabeled issues, we analyzed the performance of six representative IR-based bug localization techniques, VSM, BugLocator, BLUiR, BRTracer, DreamLoc, and GenLoc, using the original, test-cleaned, and fully-cleaned configurations of the Ye et al. dataset. Our results show that removing mislabeled issues from evaluation data consistently reduces the measured performance of all six techniques, with Accuracy@1 reductions ranging from 5.40\% to 29.26\% and statistically significant MRR reductions ranging from 4.68\% to 21.74\%. Although mislabeling does not change the overall ordering of techniques or the statistical significance of pairwise comparisons, it influences the magnitude of their performance differences. This highlights the importance of dataset curation in IRBL research to ensure fair evaluation and development of reliable debugging tools.

In the future, we will examine the impact of mislabeled issues on more fine-grained localization approaches, such as method-level \cite{youm2017improved,zhang2019finelocator,cao2020bugpecker} and change-set-based \cite{wen2016locus,ciborowska2022fast,du2023pre} techniques. Additionally, we will analyze projects written in other programming languages to assess whether the observed impact of mislabeled issues remains consistent across different development ecosystems.

% In the future, we will extend our evaluation to additional datasets to generalize our findings. Furthermore, we will examine the impact of mislabeled issues on more fine-grained approaches, such as method-level \cite{youm2017improved,zhang2019finelocator,cao2020bugpecker} and change-set-based \cite{wen2016locus,ciborowska2022fast,du2023pre} bug localization techniques.

% \bibliographystyle{ACM-Reference-Format}
% \bibliography{BL}

% \begin{scriptsize}

\bibliography{mybib}

@inproceedings{lee2018bench4bl,
  title={Bench4bl: reproducibility study on the performance of ir-based bug localization},
  author={Lee, Jaekwon and Kim, Dongsun and Bissyand{\'e}, Tegawend{\'e} F and Jung, Woosung and Le Traon, Yves},
  booktitle={ISSTA},
  pages={61--72},
  year={2018}
}

@book{wong2023handbook,
  title={Handbook of software fault localization: foundations and advances},
  author={Wong, W Eric and Tse, TH},
  year={2023},
  publisher={John Wiley \& Sons}
}

@inproceedings{raychaudhuri2008introduction,
  title={Introduction to monte carlo simulation},
  author={Raychaudhuri, Samik},
  booktitle={WSC},
  pages={91--100},
  year={2008},
}

@article{herbold2020feasibility,
  title={On the feasibility of automated prediction of bug and non-bug issues},
  author={Herbold, Steffen and Trautsch, Alexander and Trautsch, Fabian},
  journal={Empirical Software Engineering},
  volume={25},
  number={6},
  pages={5333--5369},
  year={2020},
  publisher={Springer}
}

@inproceedings{caumartin2026reformulate,
  title={Reformulate, Retrieve, Localize: Agents for Repository-Level Bug Localization},
  author={Caumartin, Genevieve and Melo, Glaucia},
  booktitle={BoatSE},
  pages={39--45},
  year={2026}
}

@article{chen2024more,
  title={Are more llm calls all you need? towards the scaling properties of compound ai systems},
  author={Chen, Lingjiao and Davis, Jared and Hanin, Boris and Bailis, Peter and Stoica, Ion and Zaharia, Matei and Zou, James},
  journal={Advances in Neural Information Processing Systems},
  volume={37},
  pages={45767--45790},
  year={2024}
}

@article{zhou2026multi,
  title={Multi-view adaptive contrastive learning for information retrieval based fault localization},
  author={Zhou, Chunying and Xie, Xiaoyuan and Chen, Gong and He, Peng and Li, Bing},
  journal={Automated Software Engineering},
  volume={33},
  number={1},
  pages={29},
  year={2026},
  publisher={Springer}
}

@article{li2025knowledge,
  title={A knowledge enhanced large language model for bug localization},
  author={Li, Yue and Liu, Bohan and Zhang, Ting and Wang, Zhiqi and Lo, David and Yang, Lanxin and Lyu, Jun and Zhang, He},
  journal={Proceedings of the ACM on Software Engineering},
  volume={2},
  number={FSE},
  pages={1914--1936},
  year={2025},
  publisher={ACM, New York, NY, USA}
}

@inproceedings{lee2024github,
  title={The github recent bugs dataset for evaluating llm-based debugging applications},
  author={Lee, Jae Yong and Kang, Sungmin and Yoon, Juyeon and Yoo, Shin},
  booktitle={ICST},
  pages={442--444},
  year={2024},
  organization={IEEE}
}

@article{asad2026towards,
  title={Towards explorative IRBL: Combining semantic retrieval with llm-driven iterative code exploration},
  author={Asad, Moumita and Yasir, Rafed Muhammad and Malek, Sam},
  journal={Proceedings of the ACM on Software Engineering},
  volume={3},
  number={ISSTA},
  pages={1376--1400},
  year={2026},
  publisher={ACM, New York, NY, USA}
}

@article{chakraborty2025blaze,
  title={BLAZE: Cross-language and cross-project bug localization via dynamic chunking and hard example learning},
  author={Chakraborty, Partha and Alfadel, Mahmoud and Nagappan, Meiyappan},
  journal={IEEE Transactions on Software Engineering},
  volume={51},
  number={8},
  pages={2254--2267},
  year={2025},
  publisher={IEEE}
}

@misc{wasserstein2016asa,
  title={The ASA statement on p-values: context, process, and purpose},
  author={Wasserstein, Ronald L and Lazar, Nicole A},
  journal={The American Statistician},
  volume={70},
  number={2},
  pages={129--133},
  year={2016},
  publisher={Taylor \& Francis}
}

@article{marzi2024k,
  title={K-Alpha calculator--krippendorff's alpha calculator: a user-friendly tool for computing krippendorff's alpha inter-rater reliability coefficient},
  author={Marzi, Giacomo and Balzano, Marco and Marchiori, Davide},
  journal={MethodsX},
  volume={12},
  pages={102545},
  year={2024},
  publisher={Elsevier}
}

@inproceedings{nashid2023retrieval,
  title={Retrieval-based prompt selection for code-related few-shot learning},
  author={Nashid, Noor and Sintaha, Mifta and Mesbah, Ali},
  booktitle={ICSE},
  pages={2450--2462},
  year={2023},
  organization={IEEE}
}

@article{7390282,
  title={A survey on software fault localization},
  author={Wong, W Eric and Gao, Ruizhi and Li, Yihao and Abreu, Rui and Wotawa, Franz},
  journal={IEEE Transactions on Software Engineering},
  volume={42},
  number={8},
  pages={707--740},
  year={2016},
  publisher={IEEE}
}

@article{colavito2025benchmarking,
  title={Benchmarking large language models for automated labeling: The case of issue report classification},
  author={Colavito, Giuseppe and Lanubile, Filippo and Novielli, Nicole},
  journal={Information and Software Technology},
  volume={184},
  pages={107758},
  year={2025},
  publisher={Elsevier}
}

@article{kang2024quantitative,
  title={A quantitative and qualitative evaluation of LLM-based explainable fault localization},
  author={Kang, Sungmin and An, Gabin and Yoo, Shin},
  journal={Proceedings of the ACM on Software Engineering},
  volume={1},
  number={FSE},
  pages={1424--1446},
  year={2024},
  publisher={ACM, New York, NY, USA}
}

@inproceedings{kong2023better,
  title={Better zero-shot reasoning with role-play prompting},
  author={Kong, Aobo and Zhao, Shiwan and Chen, Hao and Li, Qicheng and Qin, Yong and Sun, Ruiqi and Zhou, Xin and Wang, Enzhi and Dong, Xiaohang},
  booktitle={NAACL-HLT},
  pages={4099--4113},
  year={2024}
}

@article{qi2021dreamloc,
  title={Dreamloc: A deep relevance matching-based framework for bug localization},
  author={Qi, Binhang and Sun, Hailong and Yuan, Wei and Zhang, Hongyu and Meng, Xiangxin},
  journal={IEEE Transactions on Reliability},
  volume={71},
  number={1},
  pages={235--249},
  year={2021},
  publisher={IEEE}
}

@article{fejzer2021tracking,
  title={Tracking buggy files: New efficient adaptive bug localization algorithm},
  author={Fejzer, Miko{\l}aj and Nar{\k{e}}bski, Jakub and Przymus, Piotr and Stencel, Krzysztof},
  journal={IEEE Transactions on Software Engineering},
  volume={48},
  number={7},
  pages={2557--2569},
  year={2021},
  publisher={IEEE}
}

@article{santos2023tag,
  title={Tag that issue: applying API-domain labels in issue tracking systems},
  author={Santos, Fabio and Vargovich, Joseph and Trinkenreich, Bianca and Santos, Italo and Penney, Jacob and Britto, Ricardo and Pimentel, Jo{\~a}o Felipe and Wiese, Igor and Steinmacher, Igor and Sarma, Anita and others},
  journal={Empirical Software Engineering},
  volume={28},
  number={5},
  pages={116},
  year={2023},
  publisher={Springer}
}

@article{ye2015mapping,
  title={Mapping bug reports to relevant files: A ranking model, a fine-grained benchmark, and feature evaluation},
  author={Ye, Xin and Bunescu, Razvan and Liu, Chang},
  journal={IEEE Transactions on Software Engineering},
  volume={42},
  number={4},
  pages={379--402},
  year={2015},
  publisher={IEEE}
}

@inproceedings{zhou2012should,
  title={Where should the bugs be fixed? more accurate information retrieval-based bug localization based on bug reports},
  author={Zhou, Jian and Zhang, Hongyu and Lo, David},
  booktitle={ICSE},
  pages={14--24},
  year={2012},
  organization={IEEE}
}

@article{malhotra2025validating,
  title={Validating multi-objective search algorithms to predict faulty classes},
  author={Malhotra, Ruchika and Singh, Monika and Khanna, Megha},
  journal={International Journal of System Assurance Engineering and Management},
  volume={16},
  number={3},
  pages={893--913},
  year={2025},
  publisher={Springer}
}

@article{liang2022modeling,
  title={Modeling function-level interactions for file-level bug localization},
  author={Liang, Hongliang and Hang, Dengji and Li, Xiangyu},
  journal={Empirical Software Engineering},
  volume={27},
  number={7},
  pages={186},
  year={2022},
  publisher={Springer}
}

@inproceedings{sisman2012incorporating,
  title={Incorporating version histories in information retrieval based bug localization},
  author={Sisman, Bunyamin and Kak, Avinash C},
  booktitle={MSR},
  pages={50--59},
  year={2012},
  organization={IEEE}
}

@inproceedings{nguyen2011topic,
  title={A topic-based approach for narrowing the search space of buggy files from a bug report},
  author={Nguyen, Anh Tuan and Nguyen, Tung Thanh and Al-Kofahi, Jafar and Nguyen, Hung Viet and Nguyen, Tien N},
  booktitle={ASE},
  pages={263--272},
  year={2011},
  organization={IEEE}
}

@article{kim2013should,
  title={Where should we fix this bug? a two-phase recommendation model},
  author={Kim, Dongsun and Tao, Yida and Kim, Sunghun and Zeller, Andreas},
  journal={IEEE Transactions on Software Engineering},
  volume={39},
  number={11},
  pages={1597--1610},
  year={2013},
  publisher={IEEE}
}

@inproceedings{moreno2014use,
  title={On the use of stack traces to improve text retrieval-based bug localization},
  author={Moreno, Laura and Treadway, John Joseph and Marcus, Andrian and Shen, Wuwei},
  booktitle={ICSME},
  pages={151--160},
  year={2014},
  organization={IEEE}
}

@inproceedings{saha2013improving,
  title={Improving bug localization using structured information retrieval},
  author={Saha, Ripon K and Lease, Matthew and Khurshid, Sarfraz and Perry, Dewayne E},
  booktitle={ASE},
  pages={345--355},
  year={2013},
  organization={IEEE}
}

@inproceedings{lukins2008source,
  title={Source code retrieval for bug localization using latent dirichlet allocation},
  author={Lukins, Stacy K and Kraft, Nicholas A and Etzkorn, Letha H},
  booktitle={WCRE},
  pages={155--164},
  year={2008},
  organization={IEEE}
}

@inproceedings{kochhar2016practitioners,
  title={Practitioners' expectations on automated fault localization},
  author={Kochhar, Pavneet Singh and Xia, Xin and Lo, David and Li, Shanping},
  booktitle={ISSTA},
  pages={165--176},
  year={2016}
}

@article{niu2025deep,
  title={When deep learning meets information retrieval-based bug localization: A survey},
  author={Niu, Feifei and Li, Chuanyi and Liu, Kui and Xia, Xin and Lo, David},
  journal={ACM Computing Surveys},
  volume={57},
  number={11},
  pages={1--41},
  year={2025},
  publisher={ACM, New York, NY}
}

@inproceedings{pradel2020scaffle,
  title={Scaffle: Bug localization on millions of files},
  author={Pradel, Michael and Murali, Vijayaraghavan and Qian, Rebecca and Machalica, Mateusz and Meijer, Erik and Chandra, Satish},
  booktitle={ISSTA},
  pages={225--236},
  year={2020}
}

@article{wu2023large,
  title={Large language models in fault localisation},
  author={Wu, Yonghao and Li, Zheng and Zhang, Jie M and Papadakis, Mike and Harman, Mark and Liu, Yong},
  journal={arXiv Preprint arXiv:2308.15276},
  year={2023}
}

@article{kim2021datasets,
  title={Are datasets for information retrieval-based bug localization techniques trustworthy? Impact analysis of bug types on IRBL},
  author={Kim, Misoo and Lee, Eunseok},
  journal={Empirical Software Engineering},
  volume={26},
  number={3},
  pages={35},
  year={2021},
  publisher={Springer}
}

@article{thomas2013impact,
  title={The impact of classifier configuration and classifier combination on bug localization},
  author={Thomas, Stephen W and Nagappan, Meiyappan and Blostein, Dorothea and Hassan, Ahmed E},
  journal={IEEE Transactions on Software Engineering},
  volume={39},
  number={10},
  pages={1427--1443},
  year={2013},
  publisher={IEEE}
}

@article{lukins2010bug,
  title={Bug localization using latent dirichlet allocation},
  author={Lukins, Stacy K and Kraft, Nicholas A and Etzkorn, Letha H},
  journal={Information and Software Technology},
  volume={52},
  number={9},
  pages={972--990},
  year={2010},
  publisher={Elsevier}
}

@inproceedings{lam2015combining,
  title={Combining deep learning with information retrieval to localize buggy files for bug reports (n)},
  author={Lam, An Ngoc and Nguyen, Anh Tuan and Nguyen, Hoan Anh and Nguyen, Tien N},
  booktitle={ASE},
  pages={476--481},
  year={2015},
  organization={IEEE}
}

@inproceedings{lam2017bug,
  title={Bug localization with combination of deep learning and information retrieval},
  author={Lam, An Ngoc and Nguyen, Anh Tuan and Nguyen, Hoan Anh and Nguyen, Tien N},
  booktitle={ICPC},
  pages={218--229},
  year={2017},
  organization={IEEE}
}

@article{koyuncu2019d,
  title={D\&c: A divide-and-conquer approach to ir-based bug localization},
  author={Koyuncu, Anil and Bissyand{\'e}, Tegawend{\'e} F and Kim, Dongsun and Liu, Kui and Klein, Jacques and Monperrus, Martin and Traon, Yves Le},
  journal={arXiv Preprint arXiv:1902.02703},
  year={2019}
}

@article{xiao2019improving,
  title={Improving bug localization with word embedding and enhanced convolutional neural networks},
  author={Xiao, Yan and Keung, Jacky and Bennin, Kwabena E and Mi, Qing},
  journal={Information and Software Technology},
  volume={105},
  pages={17--29},
  year={2019},
  publisher={Elsevier}
}

@inproceedings{zhang2020exploiting,
  title={Exploiting code knowledge graph for bug localization via bi-directional attention},
  author={Zhang, Jinglei and Xie, Rui and Ye, Wei and Zhang, Yuhan and Zhang, Shikun},
  booktitle={ICPC},
  pages={219--229},
  year={2020}
}

@article{chakraborty2024rlocator,
  title={Rlocator: Reinforcement learning for bug localization},
  author={Chakraborty, Partha and Alfadel, Mahmoud and Nagappan, Meiyappan},
  journal={IEEE Transactions on Software Engineering},
  volume={50},
  number={10},
  pages={2695--2708},
  year={2024},
  publisher={IEEE}
}

@inproceedings{cao2020bugpecker,
  title={Bugpecker: Locating faulty methods with deep learning on revision graphs},
  author={Cao, Junming and Yang, Shouliang and Jiang, Wenhui and Zeng, Hushuang and Shen, Beijun and Zhong, Hao},
  booktitle={ASE},
  pages={1214--1218},
  year={2020}
}

@inproceedings{ciborowska2022fast,
  title={Fast changeset-based bug localization with BERT},
  author={Ciborowska, Agnieszka and Damevski, Kostadin},
  booktitle={ICSE},
  pages={946--957},
  year={2022}
}

@inproceedings{du2023pre,
  title={Pre-training code representation with semantic flow graph for effective bug localization},
  author={Du, Yali and Yu, Zhongxing},
  booktitle={FSE},
  pages={579--591},
  year={2023}
}

@article{wang2020multi,
  title={Multi-dimension convolutional neural network for bug localization},
  author={Wang, Bei and Xu, Ling and Yan, Meng and Liu, Chao and Liu, Ling},
  journal={IEEE Transactions on Services Computing},
  volume={15},
  number={3},
  pages={1649--1663},
  year={2020},
  publisher={IEEE}
}

@article{xiao2023bugradar,
  title={BugRadar: Bug localization by knowledge graph link prediction},
  author={Xiao, Xi and Xiao, Renjie and Li, Qing and Lv, Jianhui and Cui, Shunyan and Liu, Qixu},
  journal={Information and Software Technology},
  volume={162},
  pages={107274},
  year={2023},
  publisher={Elsevier}
}

@article{iqbal2020determining,
  title={Determining bug prioritization using feature reduction and clustering with classification},
  author={Iqbal, Shahid and Naseem, Rashid and Jan, Salman and Alshmrany, Sami and Yasar, Muhammad and Ali, Arshad},
  journal={IEEE Access},
  volume={8},
  pages={215661--215678},
  year={2020},
  publisher={IEEE}
}

@article{luo2022improving,
  title={Improving bug localization with effective contrastive learning representation},
  author={Luo, Zhengmao and Wang, Wenyao and Cen, Caichun},
  journal={IEEE Access},
  volume={11},
  pages={32523--32533},
  year={2022},
  publisher={IEEE}
}

@inproceedings{anh2021imbalanced,
  title={An imbalanced deep learning model for bug localization},
  author={Anh, Bui Thi Mai and Luyen, Nguyen Viet},
  booktitle={APSEC Workshops},
  pages={32--40},
  year={2021},
  organization={IEEE}
}

@article{zakari2020multiple,
title = {Multiple fault localization of software programs: A systematic literature review},
journal = {Information and Software Technology},
volume = {124},
pages = {106312},
year = {2020},
issn = {0950-5849},
doi = {https://doi.org/10.1016/j.infsof.2020.106312},
url = {https://www.sciencedirect.com/science/article/pii/S0950584920300641},
author = {Abubakar Zakari and Sai Peck Lee and Rui Abreu and Babiker Hussien Ahmed and Rasheed Abubakar Rasheed},
}

@article{han2023bjxnet,
  title={bjXnet: an improved bug localization model based on code property graph and attention mechanism},
  author={Han, Jiaxuan and Huang, Cheng and Sun, Siqi and Liu, Zhonglin and Liu, Jiayong},
  journal={Automated Software Engineering},
  volume={30},
  number={1},
  pages={12},
  year={2023},
  publisher={Springer}
}

@inproceedings{wong2014boosting,
  author={Wong, Chu-Pan and Xiong, Yingfei and Zhang, Hongyu and Hao, Dan and Zhang, Lu and Mei, Hong},
  booktitle={ICSME},
  title={Boosting Bug-Report-Oriented Fault Localization with Segmentation and Stack-Trace Analysis}, 
  year={2014},
  volume={},
  number={},
  pages={181-190},
  doi={10.1109/ICSME.2014.40}}

@inproceedings{widyasari2022influence,
  title={On the influence of biases in bug localization: Evaluation and benchmark},
  author={Widyasari, Ratnadira and Haryono, Stefanus Agus and Thung, Ferdian and Shi, Jieke and Tan, Constance and Wee, Fiona and Phan, Jack and Lo, David},
  booktitle={SANER},
  pages={128--139},
  year={2022},
  organization={IEEE}
}

@inproceedings{rath2018analyzing,
  title={Analyzing requirements and traceability information to improve bug localization},
  author={Rath, Michael and Lo, David and M{\"a}der, Patrick},
  booktitle={MSR},
  pages={442--453},
  year={2018}
}

@article{youm2017improved,
  title={Improved bug localization based on code change histories and bug reports},
  author={Youm, Klaus Changsun and Ahn, June and Lee, Eunseok},
  journal={Information and Software Technology},
  volume={82},
  pages={177--192},
  year={2017},
  publisher={Elsevier}
}

@article{khatiwada2018just,
  title={Just enough semantics: An information theoretic approach for IR-based software bug localization},
  author={Khatiwada, Saket and Tushev, Miroslav and Mahmoud, Anas},
  journal={Information and Software Technology},
  volume={93},
  pages={45--57},
  year={2018},
  publisher={Elsevier}
}

@article{wang2016amalgam+,
  title={Amalgam+: Composing rich information sources for accurate bug localization},
  author={Wang, Shaowei and Lo, David},
  journal={Journal of Software: Evolution and Process},
  volume={28},
  number={10},
  pages={921--942},
  year={2016},
  publisher={Wiley Online Library}
}

@inproceedings{wen2016locus,
  title={Locus: Locating bugs from software changes},
  author={Wen, Ming and Wu, Rongxin and Cheung, Shing-Chi},
  booktitle={ASE},
  pages={262--273},
  year={2016}
}

@inproceedings{wang2014version,
  title={Version history, similar report, and structure: Putting them together for improved bug localization},
  author={Wang, Shaowei and Lo, David},
  booktitle={ICPC},
  pages={53--63},
  year={2014}
}

@inproceedings{ye2014learning,
  title={Learning to rank relevant files for bug reports using domain knowledge},
  author={Ye, Xin and Bunescu, Razvan and Liu, Chang},
  booktitle={FSE},
  pages={689--699},
  year={2014}
}

@article{xu2025flexfl,
  title={Flexfl: Flexible and effective fault localization with open-source large language models},
  author={Xu, Chuyang and Liu, Zhongxin and Ren, Xiaoxue and Zhang, Gehao and Liang, Ming and Lo, David},
  journal={IEEE Transactions on Software Engineering},
  volume={51},
  number={5},
  pages={1455--1471},
  year={2025},
  publisher={IEEE}
}

@article{qin2024fault,
  title={Fault localization from the semantic code search perspective},
  author={Qin, Yihao and Wang, Shangwen and Lei, Yan and Zhang, Zhuo and Lin, Bo and Peng, Xin and Ma, Jun and Chen, Liqian and Mao, Xiaoguang},
  journal={ACM Transactions on Software Engineering and Methodology},
  volume={35},
  number={5},
  pages={1--32},
  year={2026},
  publisher={ACM, New York, NY}
}

@article{chang2025bridging,
  title={Bridging bug localization and issue fixing: A hierarchical localization framework leveraging large language models},
  author={Chang, Jianming and Zhou, Xin and Lulu, Lulu and Lo, David and Li, Bixin},
  journal={IEEE Transactions on Software Engineering},
  year={2026},
  publisher={IEEE}
}

@article{shi2025enhancing,
  title={Enhancing LLM-based Fault Localization with a Functionality-Aware Retrieval-Augmented Generation Framework},
  author={Shi, Xinyu and Li, Zhenhao and Chen, An Ran},
  journal={arXiv Preprint arXiv:2509.20552},
  year={2025}
}

@article{hossain2024deep,
author = {Hossain, Soneya Binta and Jiang, Nan and Zhou, Qiang and Li, Xiaopeng and Chiang, Wen-Hao and Lyu, Yingjun and Nguyen, Hoan and Tripp, Omer},
title = {A Deep Dive into Large Language Models for Automated Bug Localization and Repair},
year = {2024},
issue_date = {July 2024},
publisher = {Association for Computing Machinery},
address = {New York, NY, USA},
volume = {1},
number = {FSE},
url = {https://doi.org/10.1145/3660773},
doi = {10.1145/3660773},
journal = {Proceedings of the ACM on Software Engineering},
month = jul,
articleno = {66},
numpages = {23}
}

@article{qin2025s, author = {Qin, Yihao and Wang, Shangwen and Lou, Yiling and Dong, Jinhao and Wang, Kaixin and Li, Xiaoling and Mao, Xiaoguang}, title = {SoapFL: A Standard Operating Procedure for LLM-Based Method-Level Fault Localization}, year = {2025}, issue_date = {April 2025}, publisher = {IEEE Press}, volume = {51}, number = {4}, issn = {0098-5589}, url = {https://doi.org/10.1109/TSE.2025.3543187}, doi = {10.1109/TSE.2025.3543187}, journal = {IEEE Transactions on Software Engineering}, month = apr, pages = {1173–1187}, numpages = {15} }

@article{rafi2024enhancing,
  title={Enhancing fault localization through ordered code analysis with llm agents and self-reflection},
  author={Rafi, Md Nakhla and Kim, Dong Jae and Chen, Tse-Hsun and Wang, Shaowei},
  journal={arXiv Preprint arXiv:2409.13642},
  year={2024}
}

@inproceedings{jiang2025cosil,
  title={Issue localization via llm-driven iterative code graph searching},
  author={Jiang, Zhonghao and Ren, Xiaoxue and Yan, Meng and Jiang, Wei and Li, Yong and Liu, Zhongxin},
  booktitle={ASE},
  pages={3034--3045},
  year={2025},
  organization={IEEE}
}

@inproceedings{chen2025locagent,
  title={Locagent: Graph-guided llm agents for code localization},
  author={Chen, Zhaoling and Tang, Robert and Deng, Gangda and Wu, Fang and Wu, Jialong and Jiang, Zhiwei and Prasanna, Viktor and Cohan, Arman and Wang, Xingyao},
  booktitle={ACL},
  pages={8697--8727},
  year={2025}
}

@article{ma2508tool,
  title={Tool-integrated reinforcement learning for repo deep search},
  author={Ma, Zexiong and Peng, Chao and Zeng, Qunhong and Gao, Pengfei and Zou, Yanzhen and Xie, Bing},
  journal={arXiv Preprint arXiv:2508.03012},
  year={2025}
}

@article{yu2025orcaloca,
  title={Orcaloca: An llm agent framework for software issue localization},
  author={Yu, Zhongming and Zhang, Hejia and Zhao, Yujie and Huang, Hanxian and Yao, Matrix and Ding, Ke and Zhao, Jishen},
  journal={arXiv Preprint arXiv:2502.00350},
  year={2025}
}

@inproceedings{Brain2025,
  title={Improved IR-based Bug Localization with Intelligent Relevance Feedback},
  author={Samir, Asif Mohammed and Rahman, Mohammad Masudur},
  booktitle={ICPC},
  pages={560--571},
  year={2025},
  organization={IEEE}
}

@inproceedings{ouyang2024repograph,
  title={Repograph: Enhancing ai software engineering with repository-level code graph},
  author={Ouyang, Siru and Yu, Wenhao and Ma, Kaixin and Xiao, Zilin and Zhang, Zhihan and Jia, Mengzhao and Han, Jiawei and Zhang, Hongming and Yu, Dong},
  booktitle={ICLR},
  pages={30098--30121},
  year={2025}
}

@inproceedings{wang2024openhands,
  title={Openhands: An open platform for ai software developers as generalist agents},
  author={Wang, Xingyao and Li, Boxuan and Song, Yufan and Xu, Frank F and Tang, Xiangru and Zhuge, Mingchen and Pan, Jiayi and Song, Yueqi and Li, Bowen and Singh, Jaskirat and others},
  booktitle={ICLR},
  pages={65882--65919},
  year={2025}
}

@article{yang2024swe,
  title={Swe-agent: Agent-computer interfaces enable automated software engineering},
  author={Yang, John and Jimenez, Carlos and Wettig, Alexander and Lieret, Kilian and Yao, Shunyu and Narasimhan, Karthik and Press, Ofir},
  journal={Advances in Neural Information Processing Systems},
  volume={37},
  pages={50528--50652},
  year={2024}
}

@inproceedings{li2025llm,
  title={LLM-BL: Large language models are zero-shot rankers for bug localization},
  author={Li, Zhengliang and Jiang, Zhiwei and Huang, Qiguo and Gu, Qing},
  booktitle={ICPC},
  pages={548--559},
  year={2025},
  organization={IEEE}
}

@inproceedings{ma2025alibaba,
  title={Alibaba lingmaagent: Improving automated issue resolution via comprehensive repository exploration},
  author={Ma, Yingwei and Yang, Qingping and Cao, Rongyu and Li, Binhua and Huang, Fei and Li, Yongbin},
  booktitle={FSE},
  pages={238--249},
  year={2025}
}

@article{tao2024magis,
  title={Magis: Llm-based multi-agent framework for github issue resolution},
  author={Tao, Wei and Zhou, Yucheng and Wang, Yanlin and Zhang, Wenqiang and Zhang, Hongyu and Cheng, Yu},
  journal={Advances in Neural Information Processing Systems},
  volume={37},
  pages={51963--51993},
  year={2024}
}

@article{xia2025agentless,
  title={Demystifying llm-based software engineering agents},
  author={Xia, Chunqiu Steven and Deng, Yinlin and Dunn, Soren and Zhang, Lingming},
  journal={Proceedings of the ACM on Software Engineering},
  volume={2},
  number={FSE},
  pages={801--824},
  year={2025},
  publisher={ACM, New York, NY, USA}
}

@inproceedings{li2025swe,
  title={Swe-debate: Competitive multi-agent debate for software issue resolution},
  author={Li, Han and Shi, Yuling and Lin, Shaoxin and Gu, Xiaodong and Lian, Heng and Wang, Xin and Jia, Yantao and Huang, Tao and Wang, Qianxiang},
  booktitle={ICSE},
  pages={1380--1392},
  year={2026}
}

@inproceedings{kochhar2014potential,
  title={Potential biases in bug localization: Do they matter?},
  author={Kochhar, Pavneet Singh and Tian, Yuan and Lo, David},
  booktitle={ASE},
  pages={803--814},
  year={2014}
}

@inproceedings{herzig2013s,
  title={It's not a bug, it's a feature: how misclassification impacts bug prediction},
  author={Herzig, Kim and Just, Sascha and Zeller, Andreas},
  booktitle={ICSE},
  pages={392--401},
  year={2013},
  organization={IEEE}
}

@inproceedings{le2015information,
  title={Information retrieval and spectrum based bug localization: Better together},
  author={Le, Tien-Duy B and Oentaryo, Richard J and Lo, David},
  booktitle={FSE},
  pages={579--590},
  year={2015}
}

@inproceedings{zhang2024autocoderover,
  title={Autocoderover: Autonomous program improvement},
  author={Zhang, Yuntong and Ruan, Haifeng and Fan, Zhiyu and Roychoudhury, Abhik},
  booktitle={ISSTA},
  pages={1592--1604},
  year={2024}
}

@article{arcega2021bug,
  title={Bug localization in model-based systems in the wild},
  author={Arcega, Lorena and Font, Jaime and Haugen, {\O}ystein and Cetina, Carlos},
  journal={ACM Transactions on Software Engineering and Methodology},
  volume={31},
  number={1},
  pages={1--32},
  year={2021},
  publisher={ACM, New York, NY}
}

@inproceedings{li2022deeplabel,
  title={Deeplabel: Automated issue classification for issue tracking systems},
  author={Li, Zhong and Pan, Minxue and Pei, Yu and Zhang, Tian and Wang, Linzhang and Li, Xuandong},
  booktitle={Internetware},
  pages={231--241},
  year={2022}
}

@misc{gpt-mini,
  title        = {GPT-4o mini},
  howpublished = {\url{https://platform.openai.com/docs/models/gpt-4o-mini}},
  note         = {Accessed: 2026-06-24},
  year         = {2026}
}

@misc{repo,
  title        = {Replication Package},
  howpublished = {\url{https://github.com/mou23/Mislabeling-in-IRBL}},
}

@article{li2022empirical,
  title={An empirical study of the effectiveness of IR-based bug localization for large-scale industrial projects},
  author={Li, Wei and Li, Qingan and Ming, Yunlong and Dai, Weijiao and Ying, Shi and Yuan, Mengting},
  journal={Empirical Software Engineering},
  volume={27},
  number={2},
  pages={47},
  year={2022},
  publisher={Springer}
}

@inproceedings{kim2022empirical,
  title={An empirical study of ir-based bug localization for deep learning-based software},
  author={Kim, Misoo and Kim, Youngkyoung and Lee, Eunseok},
  booktitle={ICST},
  pages={128--139},
  year={2022},
  organization={IEEE Computer Society}
}

@inproceedings{akbar2020large,
  title={A large-scale comparative evaluation of IR-based tools for bug localization},
  author={Akbar, Shayan A and Kak, Avinash C},
  booktitle={MSR},
  pages={21--31},
  year={2020}
}

@inproceedings{tantithamthavorn2015impact,
  title={The impact of mislabelling on the performance and interpretation of defect prediction models},
  author={Tantithamthavorn, Chakkrit and McIntosh, Shane and Hassan, Ahmed E and Ihara, Akinori and Matsumoto, Kenichi},
  booktitle={ICSE},
  volume={1},
  pages={812--823},
  year={2015},
  organization={IEEE}
}

@article{wang2018deep,
  title={Deep semantic feature learning for software defect prediction},
  author={Wang, Song and Liu, Taiyue and Nam, Jaechang and Tan, Lin},
  journal={IEEE Transactions on Software Engineering},
  volume={46},
  number={12},
  pages={1267--1293},
  year={2018},
  publisher={IEEE}
}

@inproceedings{colavito2023few,
  title={Few-shot learning for issue report classification},
  author={Colavito, Giuseppe and Lanubile, Filippo and Novielli, Nicole},
  booktitle={NLBSE},
  pages={16--19},
  year={2023},
  organization={IEEE}
}

@article{pandey2017automated,
  title={Automated classification of software issue reports using machine learning techniques: an empirical study},
  author={Pandey, Nitish and Sanyal, Debarshi Kumar and Hudait, Abir and Sen, Amitava},
  journal={Innovations in Systems and Software Engineering},
  volume={13},
  number={4},
  pages={279--297},
  year={2017},
  publisher={Springer}
}

@inproceedings{kallis2019ticket,
  title={Ticket tagger: Machine learning driven issue classification},
  author={Kallis, Rafael and Di Sorbo, Andrea and Canfora, Gerardo and Panichella, Sebastiano},
  booktitle={ICSME},
  pages={406--409},
  year={2019},
  organization={IEEE}
}

@inproceedings{colavito2024leveraging,
  title={Leveraging gpt-like llms to automate issue labeling},
  author={Colavito, Giuseppe and Lanubile, Filippo and Novielli, Nicole and Quaranta, Luigi},
  booktitle={MSR},
  pages={469--480},
  year={2024}
}

@inproceedings{kochhar2014s,
  title={It's not a bug, it's a feature: does misclassification affect bug localization?},
  author={Kochhar, Pavneet Singh and Le, Tien-Duy B and Lo, David},
  booktitle={MSR},
  pages={296--299},
  year={2014}
}

@article{razali2011power,
  title={Power comparisons of shapiro-wilk, kolmogorov-smirnov, lilliefors and anderson-darling tests},
  author={Razali, Nornadiah Mohd and Wah, Yap Bee and others},
  journal={Journal of Statistical Modeling and Analytics},
  volume={2},
  number={1},
  pages={21--33},
  year={2011}
}

@article{derrick2020parametric,
  title={Parametric and non-parametric tests for the comparison of two samples which both include paired and unpaired observations},
  author={Derrick, Ben and White, Paul and Toher, Deirdre},
  journal={Journal of Modern Applied Statistical Methods},
  volume={18},
  number={1},
  pages={9},
  year={2020}
}

@article{conover1981rank,
  title={Rank transformations as a bridge between parametric and nonparametric statistics},
  author={Conover, William J and Iman, Ronald L},
  journal={The American Statistician},
  volume={35},
  number={3},
  pages={124--129},
  year={1981},
  publisher={Taylor \& Francis}
}

@phdthesis{derrick2020statistical,
  title={Statistical inference with paired observations and independent observations in two samples},
  author={Derrick, Ben},
  year={2020},
  school={University of the West of England}
}

@inproceedings{rao2011retrieval,
  title={Retrieval from software libraries for bug localization: a comparative study of generic and composite text models},
  author={Rao, Shivani and Kak, Avinash},
  booktitle={MSR},
  pages={43--52},
  year={2011}
}

@article{kerby2014simple,
  title={The simple difference formula: An approach to teaching nonparametric correlation},
  author={Kerby, Dave S},
  journal={Comprehensive Psychology},
  volume={3},
  pages={11--IT},
  year={2014},
  publisher={SAGE Publications, Sage CA: Los Angeles, CA}
}

@article{manchala2022diversity,
  title={Diversity based imbalance learning approach for software fault prediction using machine learning models},
  author={Manchala, Pravali and Bisi, Manjubala},
  journal={Applied Soft Computing},
  volume={124},
  pages={109069},
  year={2022},
  publisher={Elsevier}
}

@inproceedings{pingclasai2013classifying,
  title={Classifying bug reports to bugs and other requests using topic modeling},
  author={Pingclasai, Natthakul and Hata, Hideaki and Matsumoto, Ken-ichi},
  booktitle={APSEC},
  volume={2},
  pages={13--18},
  year={2013},
  organization={IEEE}
}

@article{le2019automated,
  title={Automated program repair},
  author={Le Goues, Claire and Pradel, Michael and Roychoudhury, Abhik},
  journal={Communications of the ACM},
  volume={62},
  number={12},
  pages={56--65},
  year={2019},
  publisher={ACM, New York, NY, USA}
}

@inproceedings{liu2019you,
  title={You cannot fix what you cannot find! an investigation of fault localization bias in benchmarking automated program repair systems},
  author={Liu, Kui and Koyuncu, Anil and Bissyand{\'e}, Tegawend{\'e} F and Kim, Dongsun and Klein, Jacques and Le Traon, Yves},
  booktitle={ICST},
  pages={102--113},
  year={2019},
  organization={IEEE}
}

@inproceedings{rahman2018improving,
  title={Improving ir-based bug localization with context-aware query reformulation},
  author={Rahman, Mohammad Masudur and Roy, Chanchal K},
  booktitle={FSE},
  pages={621--632},
  year={2018}
}

@article{assiri2017fault,
  title={Fault localization for automated program repair: effectiveness, performance, repair correctness},
  author={Assiri, Fatmah Yousef and Bieman, James M},
  journal={Software Quality Journal},
  volume={25},
  number={1},
  pages={171--199},
  year={2017},
  publisher={Springer}
}

@article{zhang2019finelocator,
  title={FineLocator: A novel approach to method-level fine-grained bug localization by query expansion},
  author={Zhang, Wen and Li, Ziqiang and Wang, Qing and Li, Juan},
  journal={Information and Software Technology},
  volume={110},
  pages={121--135},
  year={2019},
  publisher={Elsevier}
}

@article{rosner2006wilcoxon,
  title={The Wilcoxon signed rank test for paired comparisons of clustered data},
  author={Rosner, Bernard and Glynn, Robert J and Lee, Mei-Ling T},
  journal={Biometrics},
  volume={62},
  number={1},
  pages={185--192},
  year={2006},
  publisher={Wiley Online Library}
}

@article{nachar2008mann,
  title={The Mann-Whitney U: A test for assessing whether two independent samples come from the same distribution},
  author={Nachar, Nadim and others},
  journal={Tutorials in Quantitative Methods for Psychology},
  volume={4},
  number={1},
  pages={13--20},
  year={2008}
}

@article{02462954-2341-35c6-ac32-7bc139c0d8ab,
  title={Robust tests for the equality of variances},
  author={Brown, Morton B and Forsythe, Alan B},
  journal={Journal of the American Statistical Association},
  volume={69},
  number={346},
  pages={364--367},
  year={1974},
  publisher={Taylor \& Francis}
}

@inproceedings{saha2014effectiveness,
  title={On the effectiveness of information retrieval based bug localization for c programs},
  author={Saha, Ripon K and Lawall, Julia and Khurshid, Sarfraz and Perry, Dewayne E.},
  booktitle={ICSME},
  pages={161--170},
  year={2014},
}

@article{abdi2010holm,
  title={Holm’s sequential Bonferroni procedure},
  author={Abdi, Herv{\'e}},
  journal={Encyclopedia of research design},
  volume={1},
  number={8},
  pages={1--8},
  year={2010},
  publisher={Thousand Oaks, California}
}

@article{DilshenerWY18,
  author    = {Tezcan Dilshener and
               Michel Wermelinger and
               Yijun Yu},
  title     = {Locating bugs without looking back},
  journal   = {Automated Software Engineering},
  volume    = {25},
  number    = {3},
  pages     = {383--434},
  year      = {2018},
}

@inproceedings{motwani2023better,
  title={Better automatic program repair by using bug reports and tests together},
  author={Motwani, Manish and Brun, Yuriy},
  booktitle={ICSE},
  pages={1225--1237},
  year={2023},
  organization={IEEE}
}
\bibliographystyle{ieeetr}

\end{document}